\documentclass[11pt,italian,a4widepaper]{article}
\usepackage{graphics,graphicx}
\usepackage{amssymb}
\usepackage{amsmath}
\usepackage{graphicx}
\usepackage{enumerate}
\usepackage{tabularx}
\usepackage{multirow}
\usepackage{subfigure}

\usepackage[dvips]{color}

\begin{document}

%Definitions: general
\newcommand{\singlespace}{\baselineskip=12pt
\lineskiplimit=0pt \lineskip=0pt }
\def\ds{\displaystyle}

%Definitions: equations
\newcommand{\beq}{\begin{equation}}
\newcommand{\eeq}{\end{equation}}
\newcommand{\lb}{\label}
\newcommand{\beqar}{\begin{eqnarray}}
\newcommand{\eeqar}{\end{eqnarray}}
\newcommand{\und}{\underline}
\newcommand{\diam}{\stackrel{\scriptscriptstyle \diamond}}

\newcommand{\Ehat}{\hat{E}}
\newcommand{\Ahat}{\hat{A}}
\newcommand{\khat}{\hat{k}}
\newcommand{\muhat}{\hat{\mu}}
\newcommand{\mc}{M^{\scriptscriptstyle C}}
\newcommand{\mt}{M^{\scriptscriptstyle T}}
\newcommand{\mei}{M^{\scriptscriptstyle M,EI}}
\newcommand{\mec}{M^{\scriptscriptstyle M,EC}}
\newcommand{\mw}{M^{\scriptscriptstyle W}}

\newenvironment{sistema}%
{\left\lbrace\begin{array}{@{}l@{}}}%
{\end{array}\right.}

\def\ob{{\, \underline{\otimes} \,}}
\def\scalp{\mbox{\boldmath$\, \cdot \,$}}
\def\gdp{\makebox{\raisebox{-.215ex}{$\Box$}\hspace{-.778em}$\times$}}
\def\bob{\makebox{\raisebox{-.215ex}{$\Box$}\hspace{-.73em}$\scalp$}}

\def\c{{\circ}}

\def\bA{\mbox{\boldmath${\it A}$}}
\def\ba{\mbox{\boldmath${\it a}$}}
\def\bB{\mbox{\boldmath${\it B}$}}
\def\bb{\mbox{\boldmath${\it b}$}}
\def\bC{\mbox{\boldmath${\it C}$}}
\def\bc{\mbox{\boldmath${\it c}$}}
\def\bD{\mbox{\boldmath${\it D}$}}
\def\bd{\mbox{\boldmath${\it d}$}}
\def\bE{\mbox{\boldmath${\it E}$}}
\def\be{\mbox{\boldmath${\it e}$}}
\def\bF{\mbox{\boldmath${\it F}$}}
\def\bff{\mbox{\boldmath${\it f}$}}
\def\bG{\mbox{\boldmath${\it G}$}}
\def\bg{\mbox{\boldmath${\it g}$}}
\def\bH{\mbox{\boldmath${\it H}$}}
\def\bh{\mbox{\boldmath${\it h}$}}
\def\bi{\mbox{\boldmath${\it i}$}}
\def\bI{\mbox{\boldmath${\it I}$}}
\def\bj{\mbox{\boldmath${\it j}$}}
\def\bK{\mbox{\boldmath${\it K}$}}
\def\bk{\mbox{\boldmath${\it k}$}}
\def\bL{\mbox{\boldmath${\it L}$}}
\def\bl{\mbox{\boldmath${\it l}$}}
\def\bM{\mbox{\boldmath${\it M}$}}
\def\bm{\mbox{\boldmath${\it m}$}}
\def\bN{\mbox{\boldmath${\it N}$}}
\def\bn{\mbox{\boldmath${\it n}$}}
\def\b0{\mbox{\boldmath${0}$}}
\def\bo{\mbox{\boldmath${\it o}$}}
\def\bP{\mbox{\boldmath${\it P}$}}
\def\bp{\mbox{\boldmath${\it p}$}}
\def\bQ{\mbox{\boldmath${\it Q}$}}
\def\bq{\mbox{\boldmath${\it q}$}}
\def\br{\mbox{\boldmath${\it r}$}}
\def\bR{\mbox{\boldmath${\it R}$}}
\def\bS{\mbox{\boldmath${\it S}$}}
\def\bs{\mbox{\boldmath${\it s}$}}
\def\bT{\mbox{\boldmath${\it T}$}}
\def\bt{\mbox{\boldmath${\it t}$}}
\def\bU{\mbox{\boldmath${\it U}$}}
\def\bu{\mbox{\boldmath${\it u}$}}
\def\bv{\mbox{\boldmath${\it v}$}}
\def\bV{\mbox{\boldmath${\it V}$}}
\def\bw{\mbox{\boldmath${\it w}$}}
\def\bW{\mbox{\boldmath${\it W}$}}
\def\by{\mbox{\boldmath${\it y}$}}
\def\bX{\mbox{\boldmath${\it X}$}}
\def\bx{\mbox{\boldmath${\it x}$}}

\def\bbD{\overline{\bD}}
\def\bbL{\overline{\bL}}
\def\bbW{\overline{\bW}}

\def\bbeta{\mbox{\boldmath${\beta}$}}
\def\bepsilon{\mbox{\boldmath${\epsilon}$}}
\def\bvarepsilon{\mbox{\boldmath${\varepsilon}$}}
\def\bsigma{\mbox{\boldmath${\sigma}$}}
\def\bphi{\mbox{\boldmath${w}$}}
\def\bzeta{\mbox{\boldmath${\zeta}$}}

\def\bQ{\mbox{\boldmath $Q$}}

\def\Id{\mbox{\boldmath${\it I}$}}
\def\balpha{\mbox{\boldmath${\alpha}$}}
\def\bbeta{\mbox{\boldmath${\beta}$}}
\def\bGamma{\mbox{\boldmath${\Gamma}$}}
\def\bDelta{\mbox{\boldmath${\Delta}$}}
\def\bkappa{\mbox{\boldmath $\kappa$}}
\def\btau{\mbox{\boldmath $\tau$}}
\def\bnu{\mbox{\boldmath $\nu$}}
\def\bchi{\mbox{\boldmath${\chi}$}}
\def\beta{\mbox{\boldmath${\eta}$}}
\def\bxi{\mbox{\boldmath${ \xi}$}}
\def\bXi{\mbox{\boldmath${\it  \Xi}$}}
\def\bsigma{\mbox{\boldmath${\sigma}$}}
\def\bvarsigma{\mbox{\boldmath${\varsigma}$}}
\def\bSigma{\mbox{\boldmath${\Sigma}$}}
\def\bupsilon{\mbox{\boldmath $\upsilon$}}
\def\bgamma{\mbox{\boldmath $\gamma$}}
\def\bTheta{\mbox{\boldmath $\Theta$}}

\def\tr{{\sf tr}}
\def\dev{{\sf dev}}
\def\div{{\sf div}}
\def\Div{{\sf Div}}
\def\Grad{{\sf Grad}}
\def\grad{{\sf grad}}
\def\Lin{{\sf Lin}}
\def\Orth{{\sf Orth}}
\def\Unim{{\sf Unim}}
\def\Sym{{\sf Sym}}

\def\msm{\mbox{${\mathsf m}$}}

\def\msM{\mbox{${\mathsf M}$}}
\def\msS{\mbox{${\mathsf S}$}}

\def\forA{\mathbb A}
\def\forB{\mathbb B}
\def\forC{\mathbb C}
\def\forE{\mathbb E}
\def\forL{\mathbb L}
\def\forN{\mathbb N}
\def\forR{\mathbb R}

\def\capA{\mbox{\boldmath${\mathsf A}$}}
\def\capB{\mbox{\boldmath${\mathsf B}$}}
\def\capC{\mbox{\boldmath${\mathsf C}$}}
\def\capD{\mbox{\boldmath${\mathsf D}$}}
\def\capE{\mbox{\boldmath${\mathsf E}$}}
\def\capF{\mbox{\boldmath${\mathsf F}$}}
\def\capG{\mbox{\boldmath${\mathsf G}$}}
\def\capH{\mbox{\boldmath${\mathsf H}$}}
\def\capI{\mbox{\boldmath${\mathsf I}$}}
\def\capK{\mbox{\boldmath${\mathsf K}$}}
\def\capL{\mbox{\boldmath${\mathsf L}$}}
\def\capM{\mbox{\boldmath${\mathsf M}$}}
\def\capR{\mbox{\boldmath${\mathsf R}$}}
\def\capW{\mbox{\boldmath${\mathsf W}$}}

\def\C{\mbox{\boldmath${\mathcal C}$}}
\def\E{\mbox{\boldmath${\mathcal E}$}}

\def\mA{\mbox{${\mathcal A}$}}
\def\mB{\mbox{${\mathcal B}$}}
\def\mC{\mbox{${\mathcal C}$}}
\def\mD{\mbox{${\mathcal D}$}}
\def\mE{\mbox{${\mathcal E}$}}
\def\mF{\mbox{${\mathcal F}$}}
\def\mG{\mbox{${\mathcal G}$}}
\def\mH{\mbox{${\mathcal H}$}}
\def\mI{\mbox{${\mathcal I}$}}
\def\mJ{\mbox{${\mathcal J}$}}
\def\mK{\mbox{${\mathcal K}$}}
\def\mL{\mbox{${\mathcal L}$}}
\def\mM{\mbox{${\mathcal M}$}}
\def\mQ{\mbox{${\mathcal Q}$}}
\def\mR{\mbox{${\mathcal R}$}}
\def\mS{\mbox{${\mathcal S}$}}
\def\mT{\mbox{${\mathcal T}$}}
\def\mV{\mbox{${\mathcal V}$}}
\def\mY{\mbox{${\mathcal Y}$}}
\def\mZ{\mbox{${\mathcal Z}$}}

%Definitions: journals
\def\AAM{{\it Adv. Appl. Mech. }}
\def\AMM{{\it Acta Metall. Mater. }}
\def\ARMA{{\it Arch. Rat. Mech. Analysis }}
\def\AMR{{\it Appl. Mech. Rev. }}
\def\CMAME {{\it Comput. Meth. Appl. Mech. Engrg. }}
\def\CMT {{\it Cont. Mech. and Therm.}}
\def\CRAS{{\it C. R. Acad. Sci., Paris }}
\def\EFM{{\it Eng. Fract. Mech. }}
\def\EJMA{{\it Eur.~J.~Mech.-A/Solids }}
\def\IMA{{\it IMA J. Appl. Math. }}
\def\IJES{{\it Int. J. Engng. Sci. }}
\def\IJMS{{\it Int. J. Mech. Sci. }}
\def\IJNME{{\it Int. J. Numer. Meth. Eng. }}
\def\IJNAMG{{\it Int. J. Numer. Anal. Meth. Geomech. }}
\def\IJP{{\it Int. J. Plasticity }}
\def\IJSS{{\it Int. J. Solids Struct. }}
\def\IngA{{\it {Ing. Archiv }}}
\def\JACS{{\it J. Am. Ceram. Soc. }}
\def\JAM{{\it J. Appl. Mech. }}
\def\JAP{{\it J. Appl. Phys. }}
\def\JE{{\it J. Elasticity }}
\def\JM{{\it J. de M\'ecanique }}
\def\JMPS{{\it J. Mech. Phys. Solids. }}
\def\MOM{{\it Mech. Materials }}
\def\MRC{{\it Mech. Res. Comm. }}
\def\MSE{{\it Mater. Sci. Eng. }}
\def\MMS{{\it Math. Mech. Solids }}
\def\MPCPS{{\it Math. Proc. Camb. Phil. Soc. }}
\def\PRSA{{\it Proc. R. Soc. Lond., Ser. A}}
\def\PRSL{{\it Proc. R. Soc. Lond. }}
\def\QAM{{\it Quart. Appl. Math. }}
\def\QJMAM{{\it Quart. J. Mech. Appl. Math. }}
\def\ZAMP{{\it Z. angew. Math. Phys. }}

% command jump: [[ ]]

\def\salto#1#2{%
%\left[\mbox{\hspace{-#1em}}\left[#2\right]\mbox{\hspace{-#1em}}\right]}
[\mbox{\hspace{-#1em}}[#2]\mbox{\hspace{-#1em}}]}

%%%%%%%%%%%%%%%%%%%%%%%%%%%%%%%%%%%%%%%%%%%%%%%%%%%%%%%%%%%%%%%%%%%%%%%%%%%%%%%%%%%%%%%%%

\title{\bf Mindlin second-gradient elastic properties \\ from dilute two-phase Cauchy-elastic composites\\
Part II: Higher-order constitutive properties and application cases}
\author{M. Bacca, D. Bigoni\footnote{Corresponding author}, F. Dal Corso \& D. Veber \\
Department of Civil, Environmental and Mechanical Engineering\\
University of Trento, \\ via Mesiano 77, I-38123 Trento, Italy\\
e-mail: mattia.bacca@ing.unitn.it, bigoni@unitn.it,\\
francesco.dalcorso@unitn.it, daniele.veber@unitn.it}
\date{}
\maketitle

\begin{abstract}

Starting from a Cauchy elastic composite with a dilute suspension of  randomly distributed inclusions and characterized at first-order by a certain
discrepancy tensor (see part I of the present article), it is shown that the equivalent second-gradient Mindlin elastic solid:
(i.) is positive definite only when the discrepancy tensor is negative defined; (ii.)
the non-local material symmetries are the same of the discrepancy tensor, and
(iii.) the non-local effective behaviour is affected by the shape of the RVE, which does not influence the first-order homogenized response.
Furthermore, explicit derivations of non-local parameters from heterogeneous Cauchy elastic composites are obtained in the particular cases of:
 ($a$) circular cylindrical and
spherical isotropic inclusions embedded in an isotropic matrix, ($b$) $n$-polygonal cylindrical voids in an isotropic matrix, and
($c$) circular cylindrical voids in an orthortropic matrix.

\end{abstract}


\section{Introduction}

In part I of the present study (Bacca et al., 2012), a methodology has been presented to obtain an equivalent second-order
 Mindlin elastic material (Mindlin, 1964), starting from a
dilute suspension of  randomly distributed elastic inclusions embedded in an elastic matrix,  under symmetry assumptions for both the RVE
and the inclusion.
In particular,
by imposing the vanishing of
the elastic energy mismatch $\mathcal{G}$ between the heterogeneous Cauchy elastic and the Mindlin equivalent
 materials produced by the same second-order displacement boundary condition,
the equivalent second gradient elastic (SGE) solid has been found to be defined (at first-order in the volume
fraction $f\ll1$ of the inclusion phase) by the
sixth-order tensor
\beq
\lb{sol}
\begin{split}
\capA^{eq}_{ijhlmn}&=-f \frac{\rho^2}{4}\left(
\tilde{\capC}_{ihln}\delta_{jm}+
\tilde{\capC}_{ihmn}\delta_{jl}+
\tilde{\capC}_{jhln}\delta_{im}+
\tilde{\capC}_{jhmn}\delta_{il}
\right),
\end{split}
\eeq
where $\rho$ is the radius of the sphere (or circle in 2D) of inertia of the RVE, and the discrepancy tensor
$\tilde{\capC}$ is introduced to define at the first-order in $f$ the difference  between
the local constitutive tensors for the effective material $\capC^{eq}$ and the matrix $\capC^{(1)}$, so that
\beq\label{valtari}
\capC^{eq}=\capC^{(1)}+f\tilde{\capC}.
\eeq
Note that $\capA^{eq}$ is zero  {\it either} when the inclusions are not present, $f=0$,  {\it or} when the inclusion has the same
elastic properties of the matrix, $\tilde{\capC}=\b0$.

In the present part II of our study it is shown (Section \ref{sez2})
that the nonlocal material identified via second-order  match of elastic energies through the constitutive tensor (\ref{sol}):
(i.) is positive definite if and only if the discrepancy tensor is negative defined;
(ii.) shares the same material symmetries with the discrepancy tensor (obtained as homogenized
 material at first-order);
  (iii.) is affected by the RVE shape, differently from
  the homogenized response at first-order.
Moreover, a series of examples useful in view of applications are provided in Section \ref{AppCases},
in particular, the material constants
  defining the nonlocal behaviour  are explicitly obtained for dilute suspensions of isotropic
  elastic circular cylindrical inclusions, of cylindrical voids with $n$-polygonal
  cross section and of spherical elastic inclusions embedded in an isotropic matrix,
  and for dilute suspension of cylindrical voids with circular cross section distributed in an orthortropic matrix.

\section{Some properties of the effective SGE solid}\lb{sez2}

Some properties of the effective SGE solid are obtained below from the
 definition of the effective higher-order constitutive tensor $\capA^{eq}$, eqn (\ref{sol}).

\subsection{Heterogeneous Cauchy RVE leading to positive definite equivalent SGE material}\label{positivita}

\paragraph{Statement.}

For constituents characterized by a positive definite strain energy, a positive definite equivalent
SGE material is obtained
if and only if the first-order discrepancy tensor $\tilde{\capC}$ is negative definite.

\paragraph{Proof.}

For constituents characterized by a positive definite strain energy,  the first-order homogenization always leads to a positive definite
equivalent fourth-order tensor $\capC^{eq}$, so that a positive strain energy (see eqn (9) in Part I) is stored within the equivalent
 SGE material if and only if
\beq\lb{Aposdef}
\capA^{eq}_{ijhlmn}\chi_{ijh}\chi_{lmn}>0 \qquad \forall \,\bchi \neq \b0\,\,\,
\mbox{with}\,\,\, \chi_{ijk}=\chi_{jik},
\eeq
where the summation convention over repeated indices is used henceforth.
Considering the form (\ref{sol}) of  $\capA^{eq}$ (note the \lq $-$' sign),
a positive definite equivalent SGE material is obtained when
\beq\lb{Cnegdef}
\tilde{\capC}_{ijhk}\chi_{lij}\chi_{lhk}<0 \qquad \forall \,\bchi \neq \b0\,\,\,
\mbox{with}\,\,\, \chi_{ijk}=\chi_{jik} .
\eeq
Since the discrepancy tensor has the minor symmetries, $\tilde{\capC}_{ijhk}=\tilde{\capC}_{jihk}=\tilde{\capC}_{ijkh}$,
the condition (\ref{Cnegdef}) can be written as
\beq\lb{Cnegdef2}
\tilde{\capC}_{ijhk}(\chi_{lij}+\chi_{lji})(\chi_{lhk}+\chi_{lkh})<0 \qquad \forall \bchi\neq \b0\,\,\,
\mbox{with}\,\,\, \chi_{ijk}=\chi_{jik},
\eeq
which corresponds to the negative definite condition for the
fourth-order constitutive tensor $\tilde{\capC}$,
because  $\chi_{lij}+\chi_{lji}=\b0$ if and only if $\bchi=\b0$.\footnote{
The last statement can be proven as follows. With reference to a third-order tensor $\varsigma_{ijk}$, symmetric
with respect to the first two indices ($\varsigma_{ijk} = \varsigma_{jik}$), we define the tensor $\gamma_{ijk}$ as
\beq\lb{simmetrizzami}
 \gamma_{ijk}=\varsigma_{ijk}+\varsigma_{ikj},
\eeq
resulting symmetric with respect to the last two indices ($\gamma_{ijk}=\gamma_{ikj}$). Relation (\ref{simmetrizzami}) is invertible, so that
\beq\lb{simmetrizzami2}
 \varsigma_{ijk}=\frac{\gamma_{ijk}+\gamma_{jki}-\gamma_{kij}}{2},
\eeq
and therefore $\bgamma=\b0$ if and only if $\bvarsigma=\b0$.
} $\Box$

The fact that the equivalent nonlocal material is positive definite only for \lq sufficiently compliant' inclusions
was already noted by Bigoni and Drugan (2007) for Cosserat constrained rotation material and
is related to the fact that higher-order continua are stiffer than Cauchy elastic materials
(imposing boundary conditions on displacement and on its normal derivative). This effect has
also an experimental counterpart provided by Gauthier (1982), who showed micropolar effects
for porous material, but \lq anti-micropolar' behaviour for a soft matrix containing stiff inclusions.

\subsection{Higher-order material symmetries for the equivalent SGE solid}\label{simmetrie}

\paragraph{Statement.} The higher-order material symmetries of the equivalent SGE solid
coincide with the material symmetries of the first-order discrepancy tensor $\tilde{\capC}$.

\paragraph{Proof.} A class of material symmetry corresponds to indifference of a constitutive equation with respect to application of
 a class of orthogonal transformations represented through an orthogonal tensor $\bQ$, so that
 an higher-order material symmetry for the equivalent SGE material occurs when
\beq\lb{symmetrycondition}
\capA^{eq}_{ijhlmn}=Q_{ip}Q_{jq}Q_{hr}Q_{ls}Q_{mt}Q_{nu} \capA^{eq}_{pqrstu},
\eeq
while for the first-order  discrepancy tensor when
\beq\lb{symmetrycondition2}
\tilde{\capC}_{ijhk}=Q_{ip}Q_{jq}Q_{hr}Q_{ks} \tilde{\capC}_{pqrs}.
\eeq

Considering the property of orthogonal transformations ($\bQ \bQ^T = \bI$), the solution (\ref{sol}) for  $\capA^{eq}$ and that this can be inverted as
\beq\label{simm13}
\begin{array}{lll}
\tilde{\capC}_{ihln}\delta_{jm}=&\ds-\frac{1}{f}\left[
\capA^{eq}_{ijhlmn}
+\capA^{eq}_{jhimnl}
+\capA^{eq}_{hijnlm}
-\capA^{eq}_{ijhnlm}
-\capA^{eq}_{hijlmn}\right.\\[3mm]
&\left.+\capA^{eq}_{ijhmnl}
+\capA^{eq}_{jhilmn}
-\capA^{eq}_{jhinlm}
-\capA^{eq}_{hijmnl}\right],
\end{array}
\eeq
it follows that the symmetry condition for the effective higher-order tensor $\capA^{eq}$, eqn (\ref{symmetrycondition}), is equivalent
 to that for the first-order  discrepancy tensor $\tilde{\capC}$, eqn (\ref{symmetrycondition2}).\footnote{
Note that isotropic discrepancy at first-order (namely isotropic $\tilde{\capC}$)
implies isotropy of the strain-gradient equivalent material $\capA^{eq}$. On the other hand,
it is known from a numerical example by Auffray et al. (2010)
that a Cauchy composite material with an hexagonal symmetry can yield a
nonlocal anisotropic response. Their example, not referred to a dilute suspension,
is not in direct contrast with the results presented here.}$\Box$

\subsection{Influence of the volume and shape of the RVE on the higher-order constitutive response}

In addition to the dependence on the shape of the inclusion, typical of first-order homogenization,
the representation (\ref{sol}) of  $\capA^{eq}$ shows that the higher-order constitutive response in the dilute case
depends on the volume and the shape of the RVE through its radius of inertia $\rho$. This feature distinguishes
second-order homogenization from first-order, since in the latter case
$\capC^{eq}$ in the dilute case is independent of the volume and shape of the RVE.
Therefore, two composite materials $\mathcal{M}$ and $\mathcal{N}$ differing only in the geometrical distribution of the inclusions
correspond to the same equivalent local tensor $\capC^{eq}(\mathcal{M})=\capC^{eq}(\mathcal{N})$,
but lead to a different higher-order equivalent tensor $\capA^{eq}(\mathcal{M})\neq\capA^{eq}(\mathcal{N})$.

\begin{figure*}[!htcb]
  \begin{center}
\includegraphics[width= 8 cm]{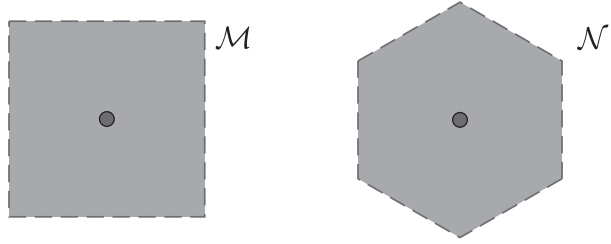}
\caption{\footnotesize Two-phase RVEs differing only in the shape of the boundary,
namely $\mathcal{M}$ and $\mathcal{N}$. In the dilute limit, both composites are characterized by the same equivalent local tensor,
$\capC^{eq}(\mathcal{M})=\capC^{eq}(\mathcal{N})$,
but by different higher-order equivalent tensors, $\capA^{eq}(\mathcal{M})\neq\capA^{eq}(\mathcal{N})$,
see eqn (\ref{tettine}).
}
 \label{ilpiuduro}
 \end{center}
\end{figure*}

\begin{figure*}[!htcb]
  \begin{center}
\includegraphics[width=9 cm]{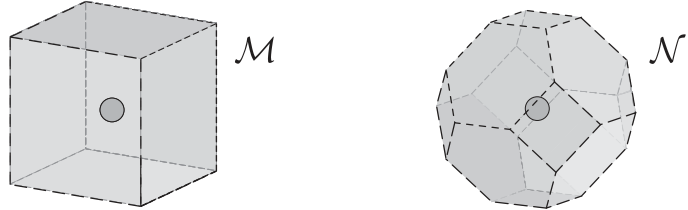}
\caption{\footnotesize Similarly to Fig. \ref{ilpiuduro},
two RVEs $\mathcal{M}$ (cubic RVE) and $\mathcal{N}$ (truncated-octahedral RVE) leading to
the same equivalent local tensor,
$\capC^{eq}(\mathcal{M})=\capC^{eq}(\mathcal{N})$,
but to different higher-order equivalent tensors, $\capA^{eq}(\mathcal{M})\neq\capA^{eq}(\mathcal{N})$,
see eqn (\ref{tettine2}).
}
 \label{ilpiuduro2}
 \end{center}
\end{figure*}

An example in 2D is reported in Fig. \ref{ilpiuduro}
where the hexagonal RVE ($\mathcal{N}$) compared to the squared RVE ($\mathcal{M}$) yields
\beq
\lb{tettine}
\capA^{eq}(\mathcal{M})=\frac{3\sqrt{3}}{5}\capA^{eq}(\mathcal{N})\sim 1.039\capA^{eq}(\mathcal{N}),
\eeq
while in the 3D example reported in Fig. \ref{ilpiuduro2} a truncated-octahedral RVE ($\mathcal{N}$)
 is compared to a cubic RVE ($\mathcal{M}$) yielding
\beq
\lb{tettine2}
\capA^{eq}(\mathcal{M})=\ds\frac{16\sqrt[3]{2}}{19}\capA^{eq}(\mathcal{N})\sim 1.061\capA^{eq}(\mathcal{N}).
\eeq
%%%%%%%%%%%%%%%%%%%%%%%%%%%%%%%%%%%%%%%%%%%%%%%%%%%%%%%%
%%%%%%%%%%%%%%%%%%%%%%%%%%%%%%%%%%%%%%%%%%%%%%%%%%%%%%%%%%
%%%%%%%%%%%%%%%%%%%%%%%%%%%%%%%%%%%%%%%%%%%%%%%%%%%%%%%%
%%%%%%%%%%%%%%%%%%%%%%%%%%%%%%%%%%%%%%%%%%%%%%%%%%%%%%%%%%

The fact that different shapes of the RVE yield, through its radius of inertia,
different nonlocal properties is inherent to the proposed identification procedure.
However, this effect is small --as shown by the estimates (\ref{tettine}) and (\ref{tettine2})-- and has to
be understood under the light of the dilute assumption for a random distribution of inclusions, so that the choice
of the shape of the RVE is to a certain extent limited.

\section{Application cases}\lb{AppCases}

Several applications of eqn (\ref{sol}) are presented in this Section for composites of different
geometries and constitutive properties. Situations in which the
homogenized material results isotropic are first considered and finally some cases of anisotropic behaviour are presented.

\subsection{Equivalent isotropic SGE}

For an isotropic composite, the first-order discrepancy tensor $\tilde{\capC}$ is
\beq
\lb{CtildeIso}
\ds\tilde{\capC}^{iso}_{ijhk}\ds=\tilde{\lambda} \delta_{ij} \delta_{hk}+\tilde{\mu} (\delta_{ih}\delta_{jk}+\delta_{ik}\delta_{jh}),
\eeq
so that the equivalent sixth-order tensor $\capA^{eq}$, eqn (\ref{sol}), is given by
\beq
\lb{aeqisotropysolution}
\begin{array}{rll}
\ds\capA^{eq}_{ijhlmn}
= &\ds-f\frac{\rho^2}{4}\left\{\tilde{\lambda}
\left[\delta_{ih}\left(\delta_{jl}\delta_{mn}+\delta_{jm}\delta_{ln}\right)
    +\delta_{jh}\left(\delta_{il}\delta_{mn}+\delta_{im}\delta_{ln}\right)\right]
\right.
\\[3mm]
\ds &\left.+\tilde{\mu}\left[
2\left(\delta_{il}\delta_{jm}+
     \delta_{im}\delta_{jl}\right)\delta_{hn}+\delta_{in}\left(\delta_{jl}\delta_{hm}+\delta_{jm}\delta_{hl}\right)
    +\delta_{jn}\left(\delta_{il}\delta_{hm}+\delta_{im}\delta_{hl}\right)\right]\right\} ,
\end{array}
\eeq
which is a special case of isotropic sixth-order tensor
\beq
\lb{isotropy_tensor}
\begin{array}{rll}
\ds\capA^{iso}_{ijhlmn}
=&\ds\frac{a_1}{2}
\left[\delta_{ij}\left(\delta_{hl}\delta_{mn}+\delta_{hm}\delta_{ln}\right)
    +\delta_{lm}\left(\delta_{in}\delta_{jh}+\delta_{ih}\delta_{jn}\right)\right]
\\[3mm]
&+\ds\frac{a_2}{2}
\left[\delta_{ih}\left(\delta_{jl}\delta_{mn}+\delta_{jm}\delta_{ln}\right)
    +\delta_{jh}\left(\delta_{il}\delta_{mn}+\delta_{im}\delta_{ln}\right)\right]
\\[3mm]
&+\ds 2\,a_3\left(\delta_{ij}\delta_{hn}\delta_{lm}\right)+a_4
\left(\delta_{il}\delta_{jm}+
     \delta_{im}\delta_{jl}\right)\delta_{hn}
\\[3mm]
&+\ds\frac{a_5}{2}
\left[\delta_{in}\left(\delta_{jl}\delta_{hm}+\delta_{jm}\delta_{hl}\right)
    +\delta_{jn}\left(\delta_{il}\delta_{hm}+\delta_{im}\delta_{hl}\right)\right] ,
\end{array}
\eeq
with the following constants
\beq\lb{solIso}
a_1=a_3=0,\qquad
a_2=-f\frac{\rho^{2}}{2}\tilde{\lambda},\qquad
a_4=a_5=-f\frac{\rho^{2}}{2}\tilde{\mu}.
\eeq

The related strain energy is positive definite
when parameters $a_i$ ($i=1,...,5$) satisfy eqn (18) of Part I, which for the values (\ref{solIso}) implies
\beq\begin{array}{c}\lb{posdefisoSol}
\tilde{K}<0,~~~
\tilde{\mu}<0,
\end{array}\eeq
where $\tilde{K}$ is the bulk modulus, equal to $\tilde{\lambda}+2\tilde{\mu}/3$ in 3D
and $\tilde{\lambda}+\tilde{\mu}$ in plane strain, and corresponding to the negative
definiteness condition for $\tilde{\capC}$, according to our previous results (Section \ref{positivita}).

An explicit evaluation of the constants ($a_{2}$, $a_{4}=a_{5}$) is given now, in the case when
an isotropic fourth-order tensor $\tilde{\capC}$ is obtained from homogenization of
a RVE with both isotropic phases, matrix denoted by \lq 1' (with Lam\'e constants $\lambda_1$ and $\mu_1$)
and inclusion denoted by \lq 2' (with Lam\'e constants $\lambda_2$ and $\mu_2$), having a shape leading to an isotropic equivalent constitutive tensor
\beq
\ds\capC_{ijhk}^{eq}\ds=\lambda_{eq} \delta_{ij} \delta_{hk}+\mu_{eq} (\delta_{ih}\delta_{jk}+\delta_{ik}\delta_{jh}),
\eeq
where
\beq
\label{legamazzo}
\lambda_{eq}=\lambda_1+f\tilde{\lambda},\qquad \mu_{eq}=\mu_1+f\tilde{\mu},\qquad K_{eq}=K_1+f\tilde{K}.
\eeq
In particular, the following forms of inclusions are considered within an isotropic matrix.
\begin{itemize}
\item For 3D deformation:
\begin{itemize}
\item spherical elastic inclusions.
\end{itemize}
\item For plane strain:
\begin{itemize}
\item circular elastic inclusions;
\item regular $n$-polygonal holes with $n \neq 4$ (the case $n=4$ leads to an orthotropic material and is treated in the next subsection).
\end{itemize}
\end{itemize}

\begin{figure*}[!htcb]
  \begin{center}
\includegraphics[width=16 cm]{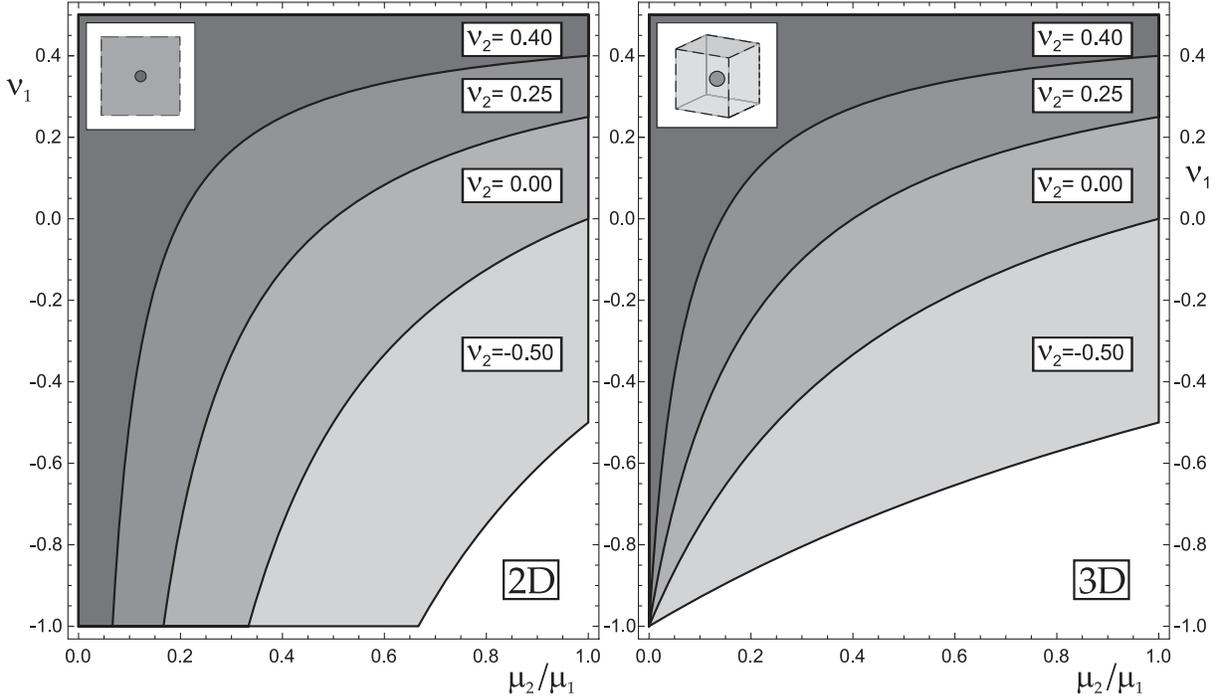}
\caption{\footnotesize Regions in the plane $\mu_2/\mu_1$ -- $\nu_1$
 where the higher-order effective constitutive tensor $\capA^{eq}$ is positive definite (for different values of $\nu_2$).
The regions for the plane strain case, eqn (\ref{posdef2d}), are reported on the left, while the case of three-dimensional
 deformations, eqn (\ref{posdef3d}), is reported on the right.}
 \label{ammissibilita}
 \end{center}
\end{figure*}

For all of the above cases it is shown that a positive definite equivalent SGE material, eqn (\ref{posdefisoSol}), is obtained only when
the inclusion phase is \lq softer' than the matrix in terms of \emph{both} shear and bulk moduli,
\beq\lb{mukappaposdef}
\mu_2<\mu_1,~~~
K_2<K_1,
\eeq
which is always satisfied when the inclusions are voids.
The positive definiteness condition (\ref{mukappaposdef}) can be written in terms of the ratio $\mu_2/\mu_1$ and the Poisson's
ratio of the phases $\nu_1$ and $\nu_2$ [where $\nu_i= \lambda_i/(2(\lambda_i+\mu_i))$] as
\beq\lb{posdef2d}
\frac{\mu_2}{\mu_1}<\min\left\{1;\,\frac{1-2\nu_2}{1-2\nu_1}\right\},
\eeq
for the case of plane strain, and
\beq\lb{posdef3d}
\frac{\mu_2}{\mu_1}<\min\left\{1;\,\frac{(1+\nu_1)(1-2\nu_2)}{(1+\nu_2)(1-2\nu_1)}\right\},
\eeq
for three-dimensional case. The regions where a positive definite SGE material is obtained,
eqns (\ref{posdef2d}) - (\ref{posdef3d}),  are mapped in the plane $\mu_2/\mu_1$ -- $\nu_1$ for
different values of the inclusion Poisson's ratio $\nu_2$
(Fig. \ref{ammissibilita}, plane strain on the left and 3D-deformation on the right).
%%%%%%%%%%%%%%%%%%%%%%%%%%%%%%%%%%%%%%%%%%%%%%%%%%%%%%%%
%%%%%%%%%%%%%%%%%%%%%%%%%%%%%%%%%%%%%%%%%%%%%%%%%%%%%%%%%%

\paragraph{Cylindrical elastic inclusions}

The elastic constants $K_{eq}$ and $\mu_{eq}$ of the
isotropic material equivalent to a dilute suspension of parallel
isotropic cylindrical inclusions embedded in an isotropic matrix have
been obtained by Hashin and Rosen (1964), in our notation
\beq
\lb{SolHes2D}
\tilde{K}=\frac{(K_2-K_1)(K_1+\mu_1)}{K_2+\mu_1},\qquad
\tilde{\mu}=\frac{2\mu_1(\mu_2-\mu_1)(K_1+\mu_1)}{2\mu_1\mu_2+K_1(\mu_1+\mu_2)}.
\eeq

\begin{figure*}[!htcb]
  \begin{center}
\includegraphics[width=16 cm]{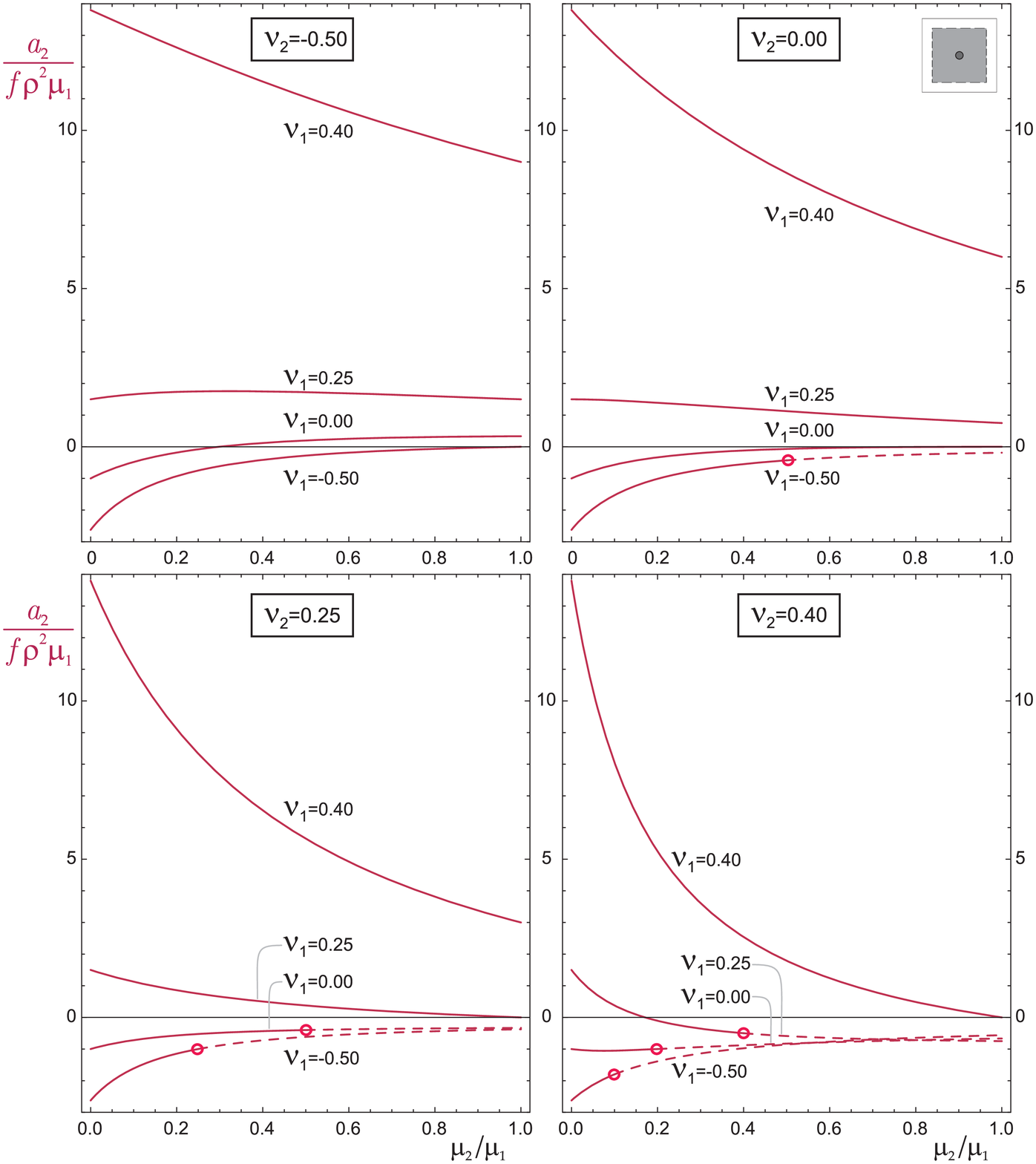}
\caption{\footnotesize Higher-order equivalent constant $a_2$, eqn (\ref{Sol2D})$_1$,
of the SGE solid equivalent to a composite made up of an isotropic matrix containing a diluite suspension of cylindrical elastic inclusions,
as a function of the ratio $\mu_2 / \mu_1$,
for different values of the Poisson's ratio of the phases $\left\{\nu_1,\nu_2\right\}=$\{-0,5;-0.25;0;0.4\}.
The constant  $a_2$ is made dimensionless
through division by parameter $f \rho^2 \mu_1$. The curves are dashed where the strain energy of
the equivalent material is not positive definite,
a red spot marks where the positive definiteness loss of $\capA^{eq}$ occurs.}
 \label{2Da2}
 \end{center}
\end{figure*}

\begin{figure*}[!htcb]
  \begin{center}
\includegraphics[width=16 cm]{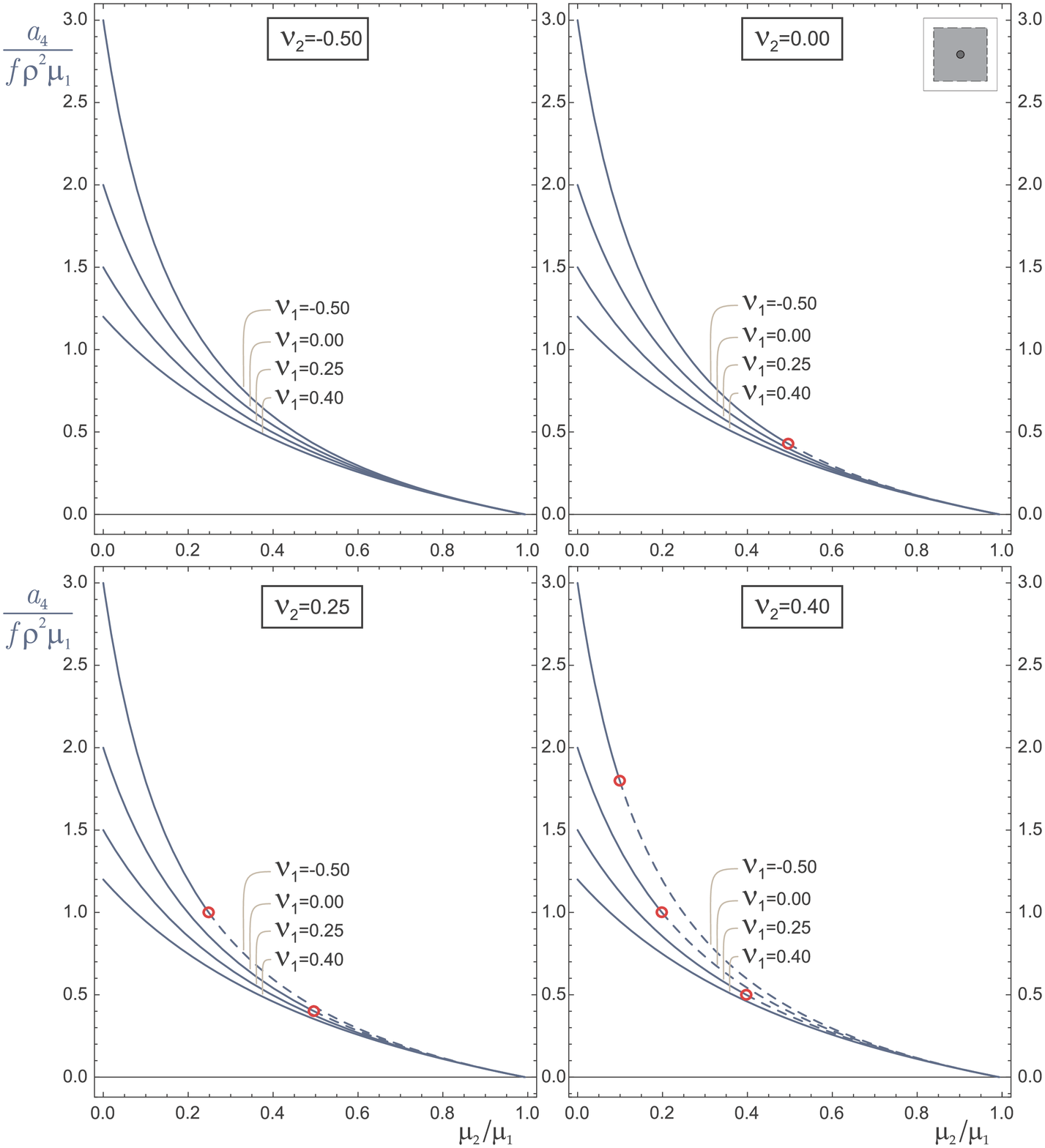}
\caption{\footnotesize Higher-order equivalent constant $a_4=a_5$, eqn (\ref{Sol2D})$_2$,
of the SGE solid equivalent to a composite made up of an isotropic matrix containing a dilute suspension of  cylindrical elastic inclusions,
as a function of the ratio $\mu_2 / \mu_1$,
for different values of Poisson's ratio of the phases $\left\{\nu_1,\nu_2\right\}=$\{-0,5;-0.25;0;0.4\}. The constant  $a_4$ is
made dimensionless
through division by parameter $f \rho^2 \mu_1$. Note that the curves are not affected by the Poisson's ratio of the
inclusion $\nu_2$, except that the threshold (red spot) for positive definiteness of the equivalent
material strain energy changes, eqn (\ref{posdef2d}).
Dashed curve represents values for which the strain energy of the equivalent material is not positive definite.}
 \label{2Da4}
 \end{center}
\end{figure*}

Exploiting equation (\ref{solIso}),
the equivalent higher-order constants  $a_i$ ($i=1,...,5$)  can be obtained from the first-order discrepancy quantities,
eqn (\ref{SolHes2D}),
so that the non-null constants are evaluated as
\beq
\lb{Sol2D}
\begin{array}{lll}
\ds a_{2}=f \frac{\rho^2}{2}\left[\frac{(K_1-K_2)(K_1+\mu_1)}{K_2+\mu_1}-\frac{\mu_1(\mu_1-\mu_2)(K_1+\mu_1)}{2\mu_1\mu_2+K_1(\mu_1+\mu_2)}\right],\\[5mm]
\ds a_{4}=a_5=f \frac{\rho^2}{2}\frac{\mu_1(\mu_1-\mu_2)(K_1+\mu_1)}{2\mu_1\mu_2+K_1(\mu_1+\mu_2)}.
\end{array}
\eeq
The higher-order equivalent constants $a_{2}$ and $a_{4}$ given by eqn (\ref{Sol2D}) are reported in Figs. \ref{2Da2} and \ref{2Da4}
as a function of the ratio $\mu_2/\mu_1$ and for different Poisson's ratios of  matrix and inclusion.
In all the figures, a red spot denotes the threshold for which the strain energy of the equivalent material looses positive definiteness.
The dashed curves refer to regions where this positive definiteness is lost.

With reference to Fig. \ref{2Da2}, we may note that $a_2\rightarrow\infty$ in the limit $\nu_1\rightarrow 1/2$. Furthermore,
$a_4$ is not affected by the Poisson's ratio of the
inclusion $\nu_2$, except that the threshold  for positive
definiteness condition for the equivalent material strain energy of the changes, eqn (\ref{posdef2d}).

%%%%%%%%%%%%%%%%%%%%%%%%%%%%%%%%%%%%%%%%%%%%%%%%%%%%%%%%
%%%%%%%%%%%%%%%%%%%%%%%%%%%%%%%%%%%%%%%%%%%%%%%%%%%%%%%%%%
%%%%%%%%%%%%%%%%%%%%%%%%%%%%%%%%%%%%%%%%%%%%%%%%%%%%%%%
%%%%%%%%%%%%%%%%%%%%%%%%%%%%%%%%%%%%%%%%%%%%%%%%%%%%%%%%%

\paragraph{Spherical elastic inclusions}

\begin{figure*}[!htcb]
  \begin{center}
\includegraphics[width=16 cm]{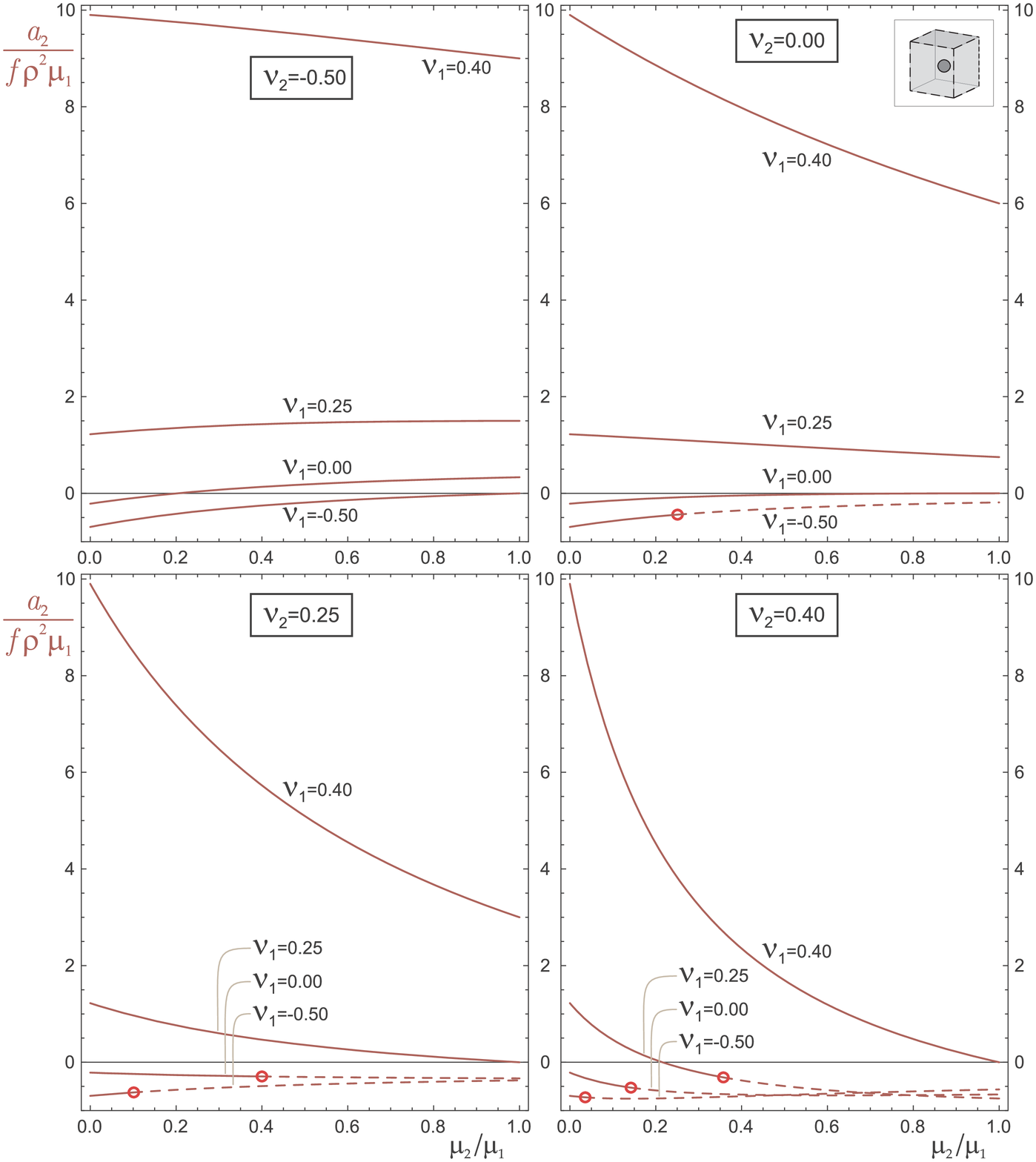}
\caption{\footnotesize Higher-order equivalent constant $a_2$, eqn (\ref{Sol3D})$_1$,
of the SGE solid equivalent to a composite made up of an isotropic matrix containing a dilute suspension of spherical elastic inclusions
as a function of the ratio $\mu_2 / \mu_1$,
for different values of Poisson's ratio of the phases $\left\{\nu_1,\nu_2\right\}=$\{-0,5;-0.25;0;0.4\}.
The constant  $a_2$ is made dimensionless
through division by parameter $f \rho^2 \mu_1$. The curves are dashed where the strain energy
of the equivalent material is not positive definite,
a red spot marks where the positive definiteness loss of $\capA^{eq}$ occurs.}
 \label{3Da2}
 \end{center}
\end{figure*}

\begin{figure*}[!htcb]
  \begin{center}
\includegraphics[width=16 cm]{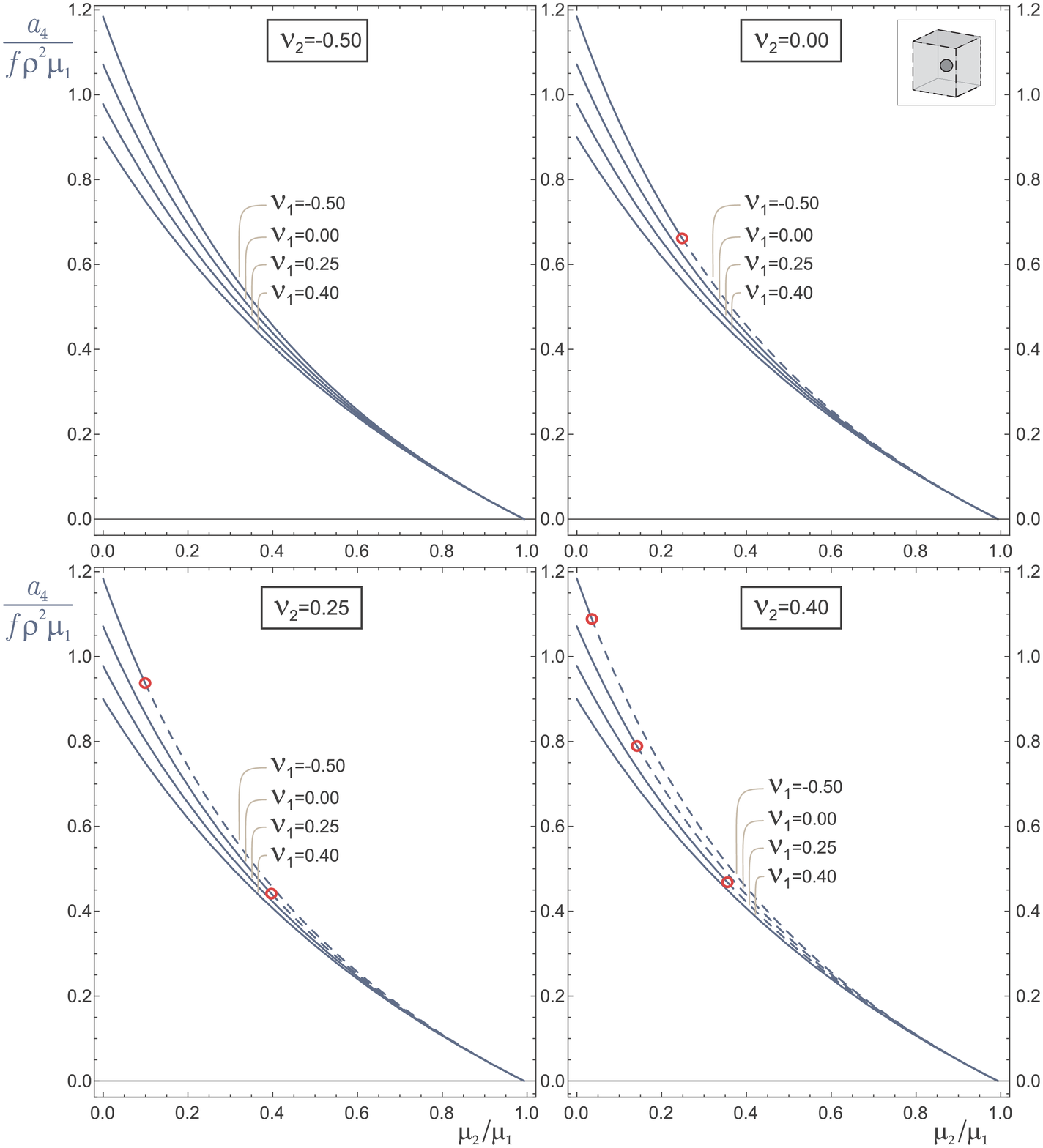}
\caption{\footnotesize Higher-order equivalent constant $a_4=a_5$, eqn (\ref{Sol3D})$_2$,
of the SGE solid equivalent to a composite made up of an isotropic matrix containing a dilute suspension of cylindrical elastic inclusions
as a function of the ratio $\mu_2 / \mu_1$,
for different values of Poisson's ratio of the phases $\left\{\nu_1,\nu_2\right\}=$\{-0,5;-0.25;0;0.4\}. The constant  $a_4$ is made dimensionless
through division by parameter $f \rho^2 \mu_1$.
Note that the curves are not affected by the Poisson's ratio of the
inclusion $\nu_2$, except that the threshold (red spot) for positive definiteness of the equivalent
material strain energy changes, eqn (\ref{posdef3d}).
Dashed curve represents values for which the strain energy of the equivalent material is not positive definite.}
 \label{3Da4}
 \end{center}
\end{figure*}

\begin{figure*}[!htcb]
  \begin{center}
\includegraphics[width=8 cm]{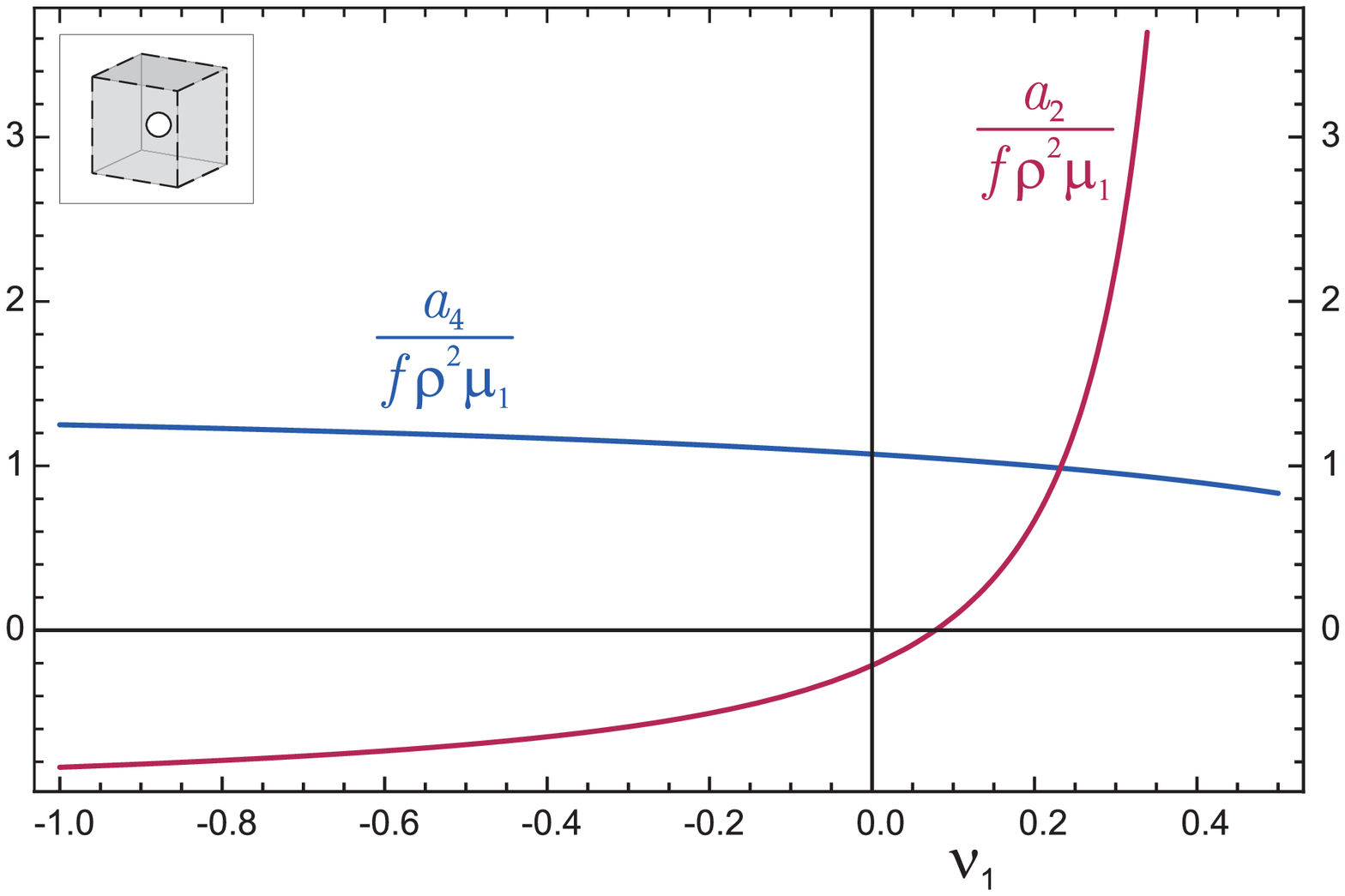}
\caption{\footnotesize Higher-order equivalent constants $a_2$ and $a_4=a_5$ of the equivalent SGE
material for a composite made up of an isotropic matrix containing a dilute suspension of spherical voids
as a function of the matrix Poisson's ratio $\nu_1$, eqn (\ref{Sol3D}) with $\mu_2=K_2=0$. The constants are made dimensionless
through division by parameter $f \rho^2 \mu_1$.}
 \label{vuoto_3D}
 \end{center}
\end{figure*}

The equivalent elastic constants $K_{eq}$ and $\mu_{eq}$ of the  isotropic
material equivalent to a dilute suspension of isotropic spherical inclusions within an isotropic matrix
have been obtained by Eshelby (1957) and independently by Hashin (1959), in our notation
\beq
\lb{SolHes3D}
\tilde{K}=\frac{(3 K_1+4 \mu_1)(K_2-K_1)}{3K_2+4 \mu_1},\qquad
\tilde{\mu}=\frac{5\mu_1(\mu_2-\mu_1)(3 K_1+4 \mu_1)}{\mu_1 (3 K_1+4 \mu_2)+2 (3 K_1+4 \mu_1) (\mu_2+\mu_1)},
\eeq
so that, through equation (\ref{solIso}),
 the non-null equivalent higher-order constants are given by
\beq
\lb{Sol3D}
\begin{array}{lll}
\ds a_{2}=f\frac{\rho^2}{2}\left[\frac{(3 K_1+4 \mu_1)(K_2-K_1)}{3K_2+4 \mu_1}-\frac{2}{3}\frac{5\mu_1(\mu_2-\mu_1)(3 K_1+4 \mu_1)}{\mu_1 (3 K_1+4 \mu_2)+2 (3
K_1+4 \mu_1) (\mu_2+\mu_1)}\right],\\[5mm]
\ds a_{4}=a_5=f\frac{\rho^2}{2}\frac{5\mu_1(\mu_2-\mu_1)(3 K_1+4 \mu_1)}{\mu_1 (3 K_1+4 \mu_2)+2 (3 K_1+4 \mu_1) (\mu_2+\mu_1)},
\end{array}
\eeq
which are reported in Fig. \ref{3Da2} and Fig. \ref{3Da4} as a function of the shear stiffness ratio $\mu_2/\mu_1$ and
for different Poisson's ratios of the phases.
In these figures the curves become dashed when the strain energy of the equivalent material looses positive definiteness.
Moreover, the higher-order constants are reported in Fig. \ref{vuoto_3D} as a function of
the matrix Poisson's ratio $\nu_1$ in the particular case of spherical voids.

Similar to the case of cylindrical elastic inclusions, $a_2\rightarrow\infty$ in the limit $\nu_1\rightarrow 1/2$ and
$a_4$ is not affected by the Poisson's ratio of the
inclusion $\nu_2$, except for the threshold of strain energy's positive definiteness, eqn. (\ref{posdef3d}).

%%%%%%%%%%%%%%%%%%%%%%%%%%%%%%%%%%%%%%%%%%%%%%%%%%%%%%%%
%%%%%%%%%%%%%%%%%%%%%%%%%%%%%%%%%%%%%%%%%%%%%%%%%%%%%%%%%%
%%%%%%%%%%%%%%%%%%%%%%%%%%%%%%%%%%%%%%%%%%%%%%%%%%%%%%%%
%%%%%%%%%%%%%%%%%%%%%%%%%%%%%%%%%%%%%%%%%%%%%%%%%%%%%%%%%%
%%%%%%%%%%%%%%%%%%%%%%%%%%%%%%%%%%%%%%%%%%%%%%%%%%%%%%%%
%%%%%%%%%%%%%%%%%%%%%%%%%%%%%%%%%%%%%%%%%%%%%%%%%%%%%%%%%%

\paragraph{Regular $n$-polygonal holes ($n\neq$4)}

\begin{figure*}[!htcb]
  \begin{center}
\includegraphics[width=8 cm]{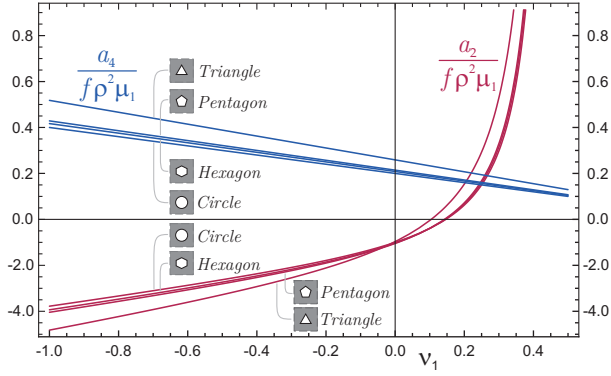}
\caption{\footnotesize
Higher-order equivalent constants $a_2$ and $a_4=a_5$ of the equivalent SGE
material for a diluite suspension of triangular ($n=3$), pentagonal ($n=5$), hexagonal ($n=6$), and circular ($n\rightarrow\infty$)
 holes in an isotropic matrix,
as functions of the matrix Poisson's ratio $\nu_1$, eqn (\ref{Sol2Dpolygon}).  The constants are made dimensionless
through division by parameter $f \rho^2 \mu_1$.}
 \label{foro_poligonale}
 \end{center}
\end{figure*}

The elastic constants $\mu_{eq}$ and $K_{eq}$ of the  isotropic
material equivalent to a dilute suspension of $n$-polygonal holes ($n\neq$4) in an isotropic matrix
have been obtained by Jasiuk et al. (1994) and Thorpe et al. (1995), from which the first-order discrepancy
stiffness can be written in our notation as
\beq
\lb{polygonal}
\tilde{K}(n)=-\mathcal{A}(n)[1-\mathcal{B}(n)]\frac{K_1+\mu_1}{\mu_1}K_1,\qquad
\tilde{\mu}(n)=-\mathcal{A}(n)[1+\mathcal{B}(n)]\frac{K_1+\mu_1}{K_1}\mu_1,
\eeq
where $\mathcal{A}(n)$ and $\mathcal{B}(n)$ are constants depending on the number of edges $n$ of the regular polygonal hole,
which can be approximated through numerical computations, and are reported in Tab. \ref{tabjasiuk}
for $n$=$\{3;5;6\}$. In the case of  a regular polygon with infinite number of edges, in other words a circle, the value of the constants is
$\mathcal{A}(n\rightarrow\infty)=3/2$
and $\mathcal{B}(n\rightarrow\infty)=1/3$, so that the case of  cylindrical void inclusion
is recovered, eqn (\ref{SolHes2D}) with $\mu_2=K_2=0$.
The equivalent higher-order constants can be obtained from eqn (\ref{solIso}) by using the first-order discrepancy quantities,
eqn (\ref{polygonal}), from which the non-null constants follow
\beq
\lb{Sol2Dpolygon}
\begin{array}{lll}
\ds a_{2}=f \frac{\rho^2}{2}\mathcal{A}(n)\left\{[1-\mathcal{B}(n)]K_1^2-[1+\mathcal{B}(n)]\mu_1^2\right\}\frac{K_1+\mu_1}{\mu_1 K_1},\\[5mm]
\ds a_{4}=a_5=f \frac{\rho^2}{2}\mathcal{A}(n)[1+\mathcal{B}(n)]\frac{K_1+\mu_1}{K_1}\mu_1,
\end{array}
\eeq
and are shown in Fig. \ref{foro_poligonale} as functions of the matrix Poisson's ratio $\nu_1$.
%%%%%%%%%%%%%%%%%%%%%%%%%%%%%%%%%%%%%%%%%%%%%%%%%%%%%%%%%%%%%%%%%%%%%%
\begin{table}[!htcb]
\centering
\renewcommand{\tablename}{\footnotesize{Tab.}}
    \begin{small}
    \begin{tabular}{|c|c|c|c|c|c|c|}
    \hline
                    &  & \multicolumn{2}{c|}{Approximated values}   \\
    \cline{3-4}
    Polygonal hole                            & $n$          & $\mathcal{A}(n)$      & $\mathcal{B}(n)$  \\
    \hline
    Triangle                         & $3$          & $2.1065$    & $0.2295$    \\
    Pentagon                         & $5$          & $1.6198$    & $0.3233$    \\
    Hexagon                          & $6$          & $1.5688$    & $0.3288$    \\
    Circle                           & $\infty$     & $3/2$         & $1/3$    \\
    \hline
    \end{tabular}
    \end{small}
\caption{\footnotesize{Values of the constants $\mathcal{A}(n)$ and
$\mathcal{B}(n)$
for triangular ($n=3$), pentagonal ($n=5$), hexagonal ($n=6$), and circular ($n\rightarrow\infty$)
 holes in an isotropic elastic matrix (Thorpe et al., 1995). These values are instrumental to obtain
 the equivalent properties $\tilde{K}(n)$ and $\tilde{\mu}(n)$, eqn (\ref{polygonal}), of the higher-order material.}}
\label{tabjasiuk}
\end{table}

%%%%%%%%%%%%%%%%%%%%%%%%%%%%%%%%%%%%%%%%%%%%%%%%%%%%%%%%
%%%%%%%%%%%%%%%%%%%%%%%%%%%%%%%%%%%%%%%%%%%%%%%%%%%%%%%%%%

\subsection{Equivalent cubic SGE}

When the first-order discrepancy tensor $\tilde{\capC}$ has a cubic symmetry,
it can be represented in a cartesian system aligned parallel to the symmetry axes as
(Thomas, 1966)
\newpage
\beq\label{cubicazzo}
\begin{array}{lll}
\ds\tilde{\capC}^{cub}_{ijhk}\ds=&\tilde{\capC}^{iso}_{ijhk}
+\tilde{\xi}\left[\left(\delta_{i2}\delta_{j3}+\delta_{i3}\delta_{j2}\right)
\left(\delta_{h2}\delta_{k3}+\delta_{h3}\delta_{k2}\right)
+\left(\delta_{i1}\delta_{j3}+\delta_{i3}\delta_{j1}\right)
\left(\delta_{h1}\delta_{k3}+\delta_{h3}\delta_{k1}\right)\right.
\\[4mm]
&\left.+\left(\delta_{i1}\delta_{j2}+\delta_{i2}\delta_{j1}\right)
\left(\delta_{h1}\delta_{k2}+\delta_{h2}\delta_{k1}\right)\right],
\end{array}
\eeq
where $\tilde{\capC}^{iso}$ is given by eqn (\ref{CtildeIso}).
The sixth-order tensor $\capA^{eq}$ for the equivalent material is obtained
using eqn (\ref{sol}) in the form
\beq \lb{aeqcubicsolution}
\begin{array}{llll}
\ds\capA^{eq}_{ijhlmn} = & \ds \capA^{iso}_{ijhlmn} +\frac{a_6}{2}
\left\{\left(\delta_{i1}\delta_{h2}+\delta_{i2}\delta_{h1}\right)\left[
\left(\delta_{l1}\delta_{n2}+\delta_{l2}\delta_{n1}\right)\delta_{jm}
+\left(\delta_{m1}\delta_{n2}+\delta_{m2}\delta_{n1}\right)\delta_{jl}
\right]\right.\\[3mm]&\ds\left.
+\left(\delta_{j1}\delta_{h2}+\delta_{j2}\delta_{h1}\right)\left[
\left(\delta_{l1}\delta_{n2}+\delta_{l2}\delta_{n1}\right)\delta_{im}
+\left(\delta_{m1}\delta_{n2}+\delta_{m2}\delta_{n1}\right)\delta_{il}
\right]\right\}
\\[3mm]
&\ds
+\left(\delta_{i1}\delta_{h3}+\delta_{i3}\delta_{h1}\right)\left[
\left(\delta_{l1}\delta_{n3}+\delta_{l3}\delta_{n1}\right)\delta_{jm}
+\left(\delta_{m1}\delta_{n3}+\delta_{m3}\delta_{n1}\right)\delta_{jl}
\right]\\[3mm]
&\ds
+\left(\delta_{j1}\delta_{h3}+\delta_{j3}\delta_{h1}\right)\left[
\left(\delta_{l1}\delta_{n3}+\delta_{l3}\delta_{n1}\right)\delta_{im}
+\left(\delta_{m1}\delta_{n3}+\delta_{m3}\delta_{n1}\right)\delta_{il}
\right]
\\[3mm]
&+\ds
\left(\delta_{i2}\delta_{h3}+\delta_{i3}\delta_{h2}\right)\left[
\left(\delta_{l2}\delta_{n3}+\delta_{l3}\delta_{n2}\right)\delta_{jm}
+\left(\delta_{m2}\delta_{n3}+\delta_{m3}\delta_{n2}\right)\delta_{jl}
\right]\\[3mm]
&\ds\left.
+\left(\delta_{j2}\delta_{h3}+\delta_{j3}\delta_{h2}\right)\left[
\left(\delta_{l2}\delta_{n3}+\delta_{l3}\delta_{n2}\right)\delta_{im}
+\left(\delta_{m2}\delta_{n3}+\delta_{m3}\delta_{n2}\right)\delta_{il}
\right]\right\},
\end{array}
\eeq
with $\capA^{iso}$ given by eqn (\ref{isotropy_tensor}), parameters $a_i\;(i=1,...,5)$ by eqn (\ref{solIso}), and
\beq\lb{solCub}
a_6=-f\frac{\rho^2}{2}\tilde{\xi}.
\eeq

According to results presented in subsections \ref{positivita} and \ref{simmetrie},
the effective higher-order tensor $\capA^{eq}$ results to be a cubic sixth-order tensor
and is positive definite when $\tilde{\capC}$, eqn (\ref{cubicazzo}), is negative
definite, namely, eqn (\ref{posdefisoSol}) together with
\beq\begin{array}{c}\lb{posdefcubSol}
\tilde{\xi}+\tilde{\mu}<0.
\end{array}\eeq

\paragraph{Aligned square holes within an isotropic matrix}

There are no results available for the plane strain homogenization of a dilute suspension
of square holes periodically distributed (with parallel edges) within an isotropic matrix.
Therefore, we have compared with a conformal mapping technique (Misseroni et al. 2013)
stress and strain averages, and found the following discrepancy at
first-order in the constitutive quantities\footnote{
Thorpe et
al. (1995) give results for composites with a random orientation of square holes,
so that the effective behaviour is isotropic and given
by eqn (\ref{polygonal}) with $\mathcal{A}(n=4)=1.738$ and
$\mathcal{B}(n=4)=0.306$. This isotropic effective response can be
independently obtained by averaging the cubic effective response
given by eqn (\ref{square}) over two orientations of the square hole
differing by an angle $\pi/4$.
}
\beq
\lb{square}
\tilde{\lambda}=-(1.198 K_1^2 - 1.864 \mu_1^2)\frac{K_1+\mu_1}{K_1\mu_1},\qquad
\tilde{\mu}=-1.864\frac{K_1+\mu_1}{K_1}\mu_1,\qquad
\tilde{\xi}=-0.796\frac{K_1+\mu_1}{K_1}\mu_1,
\eeq
showing that
$\tilde{\capC}$ is negative definite, eqn (\ref{posdefcubSol}), and
therefore the corresponding effective higher-order tensor
$\capA^{eq}$, eqn (\ref{aeqcubicsolution}), is positive definite.

The equivalent higher-order constants $a_i$ ($i=1,...,6$) can be obtained from the first-order discrepancy quantities,
eqn (\ref{square}), so that the non-null constants
are evaluated by exploiting eqns (\ref{solIso}) and (\ref{solCub}) as
\beq
\lb{Sol2Dsquare}
\begin{array}{lll}
\ds a_{2}=f \rho^2\,(0.599 K_1^2 - 0.932 \mu_1^2)\frac{K_1+\mu_1}{K_1\mu_1},\\[5mm]
\ds a_{4}=a_5= 0.932 f \rho^2\frac{K_1+\mu_1}{K_1}\mu_1,\\[5mm]
\ds a_{6}=0.398 f \rho^2\frac{K_1+\mu_1}{K_1}\mu_1.
\end{array}
\eeq

\begin{figure*}[!htcb]
  \begin{center}
\includegraphics[width=8 cm]{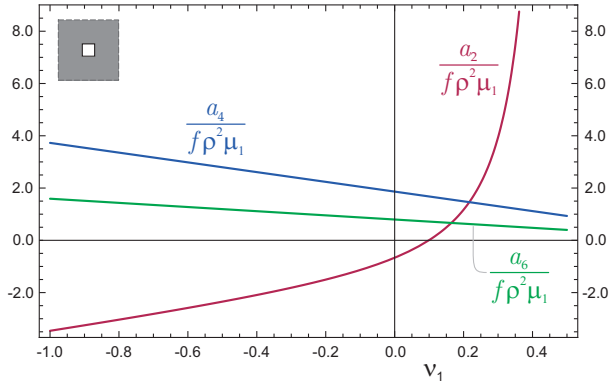}
\caption{\footnotesize Higher-order equivalent constants $a_2$, $a_4=a_5$, and $a_6$ of the equivalent SGE
material for the plane strain case of a dilute suspension of periodically-distributed (with parallel edges) square holes in an isotropic matrix,
as a function of the matrix Poisson's ratio $\nu_1$, eqn (\ref{Sol2Dsquare}).  The constants are made dimensionless
through division by parameter $f \rho^2 \mu_1$.
}
 \label{foro_quadrato}
 \end{center}
\end{figure*}

These three independent constants are reported in  Fig. \ref{foro_quadrato} as functions of
the matrix Poisson's ratio $\nu_1$.
%%%%%%%%%%%%%%%%%%%%%%%%%%%%%%%%%%%%%%%%%%%%%%%%%%%%%%%%
%%%%%%%%%%%%%%%%%%%%%%%%%%%%%%%%%%%%%%%%%%%%%%%%%%%%%%%%%%

\subsection{Equivalent orthotropic SGE}

When the first-order discrepancy tensor
$\tilde{\capC}$ is orthotropic, it can be represented in a cartesian system aligned parallel to the symmetry axes as
(Spencer, 1982)
\beq
\label{ortopedico}
\begin{array}{lll}
\ds\tilde{\capC}^{orth}_{ijhk}\ds=&
\tilde{\capC}^{iso}_{ijhk}
+\tilde{\xi}_I\left(\delta_{i2}\delta_{j3}+\delta_{i3}\delta_{j2}\right)
\left(\delta_{h2}\delta_{k3}+\delta_{h3}\delta_{k2}\right)
+\tilde{\xi}^{II}\left(\delta_{i1}\delta_{j3}+\delta_{i3}\delta_{j1}\right)
\left(\delta_{h1}\delta_{k3}+\delta_{h3}\delta_{k1}\right)
\\[4mm]
&+\tilde{\xi}^{III}\left(\delta_{i1}\delta_{j2}+\delta_{i2}\delta_{j1}\right)
\left(\delta_{h1}\delta_{k2}+\delta_{h2}\delta_{k1}\right)
+\tilde{\omega}_I\delta_{i1}\delta_{j1}\delta_{h1}\delta_{k1}
+\tilde{\omega}^{II}\delta_{i3}\delta_{j3}\delta_{h3}\delta_{k3}
\\[4mm]
&+\tilde{\omega}^{III}\left(\delta_{ij}\delta_{h3}\delta_{k3}+\delta_{hk}\delta_{i3}\delta_{j3}\right)
+\tilde{\omega}^{IV}\left(\delta_{i1}\delta_{j1}\delta_{h3}\delta_{k3}+\delta_{i3}\delta_{j3}\delta_{h1}\delta_{k1}\right),
\end{array}
\eeq
where $\tilde{\xi}^{III}$, $\tilde{\omega}^{I}$,
$\tilde{\xi}^{I}$, $\tilde{\xi}^{II}$, $\tilde{\omega}^{II}$,
$\tilde{\omega}^{III}$ and $\tilde{\omega}^{IV}$ are seven independent
constants (in addition to $\tilde{\lambda}$ and $\tilde{\mu}$) defining the orthotropic
behaviour in 3D.\footnote{
Note that the cubic representation (\ref{cubicazzo}) is obtained as a particular case
by setting $\tilde{\xi}^{I}=\tilde{\xi}^{II}=\tilde{\xi}^{I}=\tilde{\xi}$ and
$\tilde{\omega}^{I}=\tilde{\omega}^{II}=\tilde{\omega}^{III}=\tilde{\omega}^{IV}=0$.
}
The in-plane behaviour is defined by groups of four independent constants, which for the $x_1$--$x_2$
plane are $\{\tilde{\lambda};\tilde{\mu};\tilde{\xi}^{III};\tilde{\omega}^{I}\}$.

In the case of orthotropic $\tilde{\capC}$, eqn (\ref{sol}) defining the sixth-order nonlocal tensor $\capA^{eq}$
leads to
\beq \lb{aeqorthotropicsolution}
\begin{array}{llll}
\ds\capA^{eq}_{ijhlmn} = & \ds \capA^{iso}_{ijhlmn} +\frac{a_6}{2}
\left\{\left(\delta_{i1}\delta_{h2}+\delta_{i2}\delta_{h1}\right)\left[
\left(\delta_{l1}\delta_{n2}+\delta_{l2}\delta_{n1}\right)\delta_{jm}
+\left(\delta_{m1}\delta_{n2}+\delta_{m2}\delta_{n1}\right)\delta_{jl}
\right]\right.\\[3mm]&\ds\left.
+\left(\delta_{j1}\delta_{h2}+\delta_{j2}\delta_{h1}\right)\left[
\left(\delta_{l1}\delta_{n2}+\delta_{l2}\delta_{n1}\right)\delta_{im}
+\left(\delta_{m1}\delta_{n2}+\delta_{m2}\delta_{n1}\right)\delta_{il}
\right]\right\}
\\[3mm]
&+\ds\frac{a_7}{2}
\left\{\left(\delta_{i1}\delta_{h3}+\delta_{i3}\delta_{h1}\right)\left[
\left(\delta_{l1}\delta_{n3}+\delta_{l3}\delta_{n1}\right)\delta_{jm}
+\left(\delta_{m1}\delta_{n3}+\delta_{m3}\delta_{n1}\right)\delta_{jl}
\right]\right.\\[3mm]
&\ds\left.
+\left(\delta_{j1}\delta_{h3}+\delta_{j3}\delta_{h1}\right)\left[
\left(\delta_{l1}\delta_{n3}+\delta_{l3}\delta_{n1}\right)\delta_{im}
+\left(\delta_{m1}\delta_{n3}+\delta_{m3}\delta_{n1}\right)\delta_{il}
\right]\right\}
\\[3mm]
&+\ds\frac{a_8}{2}
\left\{\left(\delta_{i2}\delta_{h3}+\delta_{i3}\delta_{h2}\right)\left[
\left(\delta_{l2}\delta_{n3}+\delta_{l3}\delta_{n2}\right)\delta_{jm}
+\left(\delta_{m2}\delta_{n3}+\delta_{m3}\delta_{n2}\right)\delta_{jl}
\right]\right.\\[3mm]
&\ds\left.
+\left(\delta_{j2}\delta_{h3}+\delta_{j3}\delta_{h2}\right)\left[
\left(\delta_{l2}\delta_{n3}+\delta_{l3}\delta_{n2}\right)\delta_{im}
+\left(\delta_{m2}\delta_{n3}+\delta_{m3}\delta_{n2}\right)\delta_{il}
\right]\right\}
\\[3mm]
&+\ds\frac{a_9}{2} \left[\delta_{i1}\left(\delta_{l1}\delta_{jm}+
\delta_{m1}\delta_{jl}\right)
+\delta_{j1}\left(\delta_{l1}\delta_{im}
+\delta_{m1}\delta_{il}\right)\right]\delta_{h1}\delta_{n1}
\\[3mm]
&+\ds\frac{a_{10}}{2} \left[\delta_{i3}\left(\delta_{l3}\delta_{jm}+
\delta_{m3}\delta_{jl}\right)
+\delta_{j3}\left(\delta_{l3}\delta_{im}
+\delta_{m3}\delta_{il}\right)\right]\delta_{h3}\delta_{n3}
\\[3mm]
&+\ds\frac{a_{11}}{2} \left\{ \delta_{h3}
\left[\delta_{ln}\left(\delta_{jm}\delta_{i3}+\delta_{im}\delta_{j3}\right)
+
\delta_{mn}\left(\delta_{jl}\delta_{i3}+\delta_{il}\delta_{j3}\right)
\right]
\right.\\[3mm]\ds&\left.
+\delta_{n3}
\left[\delta_{ih}\left(\delta_{jm}\delta_{l3}+\delta_{jl}\delta_{m3}\right)
+
\delta_{jh}\left(\delta_{im}\delta_{l3}+\delta_{il}\delta_{m3}\right)\right]\right\}\\[3mm]
&+\ds\frac{a_{12}}{2} \left\{
\delta_{h1}\delta_{n3}
\left[\delta_{i1}\left(\delta_{jm}\delta_{l3}+\delta_{jl}\delta_{m3}\right)
+
\delta_{j1}\left(\delta_{im}\delta_{l3}+\delta_{il}\delta_{m3}\right)
\right]\right.
\\[3mm]\ds&\left.
\delta_{h3}\delta_{n1}
\left[\delta_{i3}\left(\delta_{jm}\delta_{l1}+\delta_{jl}\delta_{m1}\right)
+
\delta_{j3}\left(\delta_{im}\delta_{l1}+\delta_{il}\delta_{m1}\right)
\right]
\right\},
\end{array}
\eeq
with $\capA^{iso}$ given by eqn (\ref{isotropy_tensor}), parameters $a_i\;(i=1,...,5)$ by eqn (\ref{solIso}), and
\beq\begin{array}{ccc}\lb{nonlocalortotropo}
a_{6}=\ds-f\frac{\rho^2}{2}\tilde{\xi}^{III},
\qquad a_{7}=\ds-f\frac{\rho^2}{2}\tilde{\xi}^{II},
\qquad a_{8}=\ds-f\frac{\rho^2}{2}\tilde{\xi}^{I},
\\[3mm]
a_{9}=\ds-f\frac{\rho^2}{2}\tilde{\omega}^{I},
\qquad a_{10}=\ds-f\frac{\rho^2}{2}\tilde{\omega}^{II},
\qquad a_{11}=\ds-f\frac{\rho^2}{2}\tilde{\omega}^{III},
\qquad a_{12}=\ds-f\frac{\rho^2}{2}\tilde{\omega}^{IV},
\end{array}\eeq

According to the results presented in subsections \ref{positivita} and \ref{simmetrie},
the effective higher-order tensor $\capA^{eq}$ results to be an orthotropic sixth-order tensor,
positive definite when $\tilde{\capC}$, eqn (\ref{ortopedico}), is negative definite, namely
\beq
\lb{posdeforthSol}
\left\{\begin{array}{lll}
\tilde{\mu}+\tilde{\xi}^{III}<0,\\[3mm]
\tilde{\mu}+\tilde{\xi}^{II}<0,\\[3mm]
\tilde{\mu}+\tilde{\xi}^{I}<0,\\[3mm]
\tilde{\lambda}+2\tilde{\mu}+\tilde{\omega}_I<0,\\[3mm]
4\tilde{\mu}(\tilde{\lambda}+\tilde{\mu})+(\tilde{\lambda}+2\tilde{\mu})\tilde{\omega}_I<0,\\[3mm]
8\tilde{\mu}^3-\tilde{\omega}_I \tilde{\omega}^{III\,2}+4\tilde{\mu} ^2 (\tilde{\omega}^{I}+\tilde{\omega}^{II}+2\tilde{\omega}^{III})
+\tilde{\lambda}\left(12\tilde{\mu}^2+\tilde{\omega}_I\tilde{\omega}^{II}+4\tilde{\mu}(\tilde{\omega}^{I}+\tilde{\omega}^{II}-\tilde{\omega}^{IV})
-\tilde{\omega}^{IV\,2}\right)\\[3mm]
\qquad-2\tilde{\mu}\left(2\tilde{\omega}^{III\,2}-\tilde{\omega}_I (\tilde{\omega}^{II}+2\tilde{\omega}^{III})+2\tilde{\omega}^{III}
\tilde{\omega}^{IV}+\tilde{\omega}^{IV\,2}\right)<0,
\end{array}\right.\eeq
while in the case of plane strain, conditions (\ref{posdeforthSol}) become, in the $x_1$--$x_2$ plane
\beq\left\{\begin{array}{lll}
\lb{posdeforthSol2}
\tilde{\mu}+\tilde{\xi}^{III}<0,\\[3mm]
\tilde{\lambda}+2\tilde{\mu}+\tilde{\omega}_I<0,\\[3mm]
4\tilde{\mu}(\tilde{\lambda}+\tilde{\mu})+(\tilde{\lambda}+2\tilde{\mu})\tilde{\omega}_I<0.
\end{array}\right.
\eeq

\paragraph{Orthotropic matrix with cylindrical holes}

We consider the plane strain of an orthotropic matrix containing a dilute suspension of circular holes with centers
aligned parallel to the orthotropy symmetry axes. In particular, assuming $x_3$ as the out-of-plane direction and
$x_1$ and $x_2$ as the orthotropy axes,
the discrepancy tensor has the form (\ref{ortopedico}) and is characterized by the following constants
\footnote{
For conciseness, in this subsection  the in-plane orthotropy parameters $\xi^{III}$ and $\omega^{I}$ are denoted
by $\xi$ and $\omega$, respectively, in the representation of both matrix and discrepancy quantities.
}
(Tsukrov  and Kachanov, 2000)
\beq
\begin{array}{lll}
\tilde{\lambda}
= & \ds
\frac{\gamma (\lambda_{1} +2 \mu_{1} ) \left\{\left[(-1+\gamma)^2-(1+\gamma)
\delta\right]\lambda_{1} ^2+2 \left[2 (-1+\gamma) \gamma-(1+\gamma) \delta\right]
\lambda_{1} \mu_{1} +4 \gamma^2 \mu_{1} ^2\right\}}{\left[(-1+\gamma) \lambda_{1}+2 \gamma \mu_{1}\right]
(\lambda_{1} +\gamma \lambda_{1} +2 \gamma \mu_{1} )},
\\[8mm]
\tilde{\mu}
= &
-\left(\lambda_{1} +2 \mu_{1} \right) \times
\\[4mm] &\ds
\frac{\left(-1+\gamma^2\right)
(-1+\gamma-\delta) \lambda_{1} ^2+2 (-1+\gamma)
\gamma (2+2\gamma-\delta) \lambda_{1}\mu_{1}
+4 \gamma \left(\gamma+\gamma^2+\delta\right)
\mu_{1} ^2}{2 \left[(-1+\gamma) \lambda_{1} +2\gamma\mu_{1} \right] (\lambda_{1} +\gamma
\lambda_{1} +2 \gamma \mu_{1})},
\\[8mm]
\tilde{\xi}
= & \ds -\tilde{\mu}
-\frac{\delta (1+\gamma+\delta)(\lambda_{1} +2 \mu_{1} )
\left[(-1+\gamma) \lambda_{1} +2 \gamma \mu_{1} \right] (\lambda_{1}+\gamma
\lambda_{1} +2 \gamma \mu_{1})}{\left[\left(-2+2 \gamma-\delta^2\right)
\lambda_{1} +4 \gamma\mu_{1} -2 \delta^2 \mu_{1} \right]^2}.
\\[8mm]
\tilde{\omega}
= & \ds -\tilde{\mu}
-\gamma \left(\lambda_{1} +2 \mu_{1} \right)\times
\\[4mm] &\ds
\frac{\left(-1+\gamma^2\right)
(-1+\gamma+\gamma \delta)\lambda_{1} ^2+2 (-1+\gamma)
\left[\delta+2\gamma (1+\gamma)
 (1+\delta)\right] \lambda_{1}\mu_{1}+4\gamma^2 (1+\gamma+\gamma \delta) \mu_{1}^2}{2 \left[(-1+\gamma) \lambda_{1} +2 \gamma\mu_{1} \right] (\lambda_{1}+\gamma \lambda_{1} +2 \gamma \mu_{1} )},
\end{array}\eeq
where
\beq
\begin{array}{ccc}
\ds\gamma=\sqrt{\Gamma^2-\Delta},
\qquad
\ds\delta=\sqrt{\Gamma+\sqrt{\Delta}}+\sqrt{\Gamma-\sqrt{\Delta}},
\qquad
\ds\Gamma\ds=\frac{2 \mu_1  (\mu_1 +\omega_1)+\lambda_1  (\mu_1 -\xi_1+\omega_1)}{(\lambda_1 +2 \mu_1 ) (\mu_1 +\xi_1)},
\\[6mm]
\ds\Delta\ds=\frac{[-2\xi_1 (\lambda_1 +2 \mu_1 +\xi_1)+(\lambda_1 +2 \mu_1 ) \omega_1] [2 \mu_1  (\mu_1 +\omega_1)
+\lambda_1 (2 \mu_1 +\omega_1)]}{(\lambda_1 +2 \mu_1 )^2 (\mu_1 +\xi_1)^2}.
\end{array}\eeq

The non-null constants $a_2$, $a_4=a_5$, $a_6$, and $a_9$ defining the effective higher-order  tensor $\capA^{eq}$
can explicitly be evaluated using eqns (\ref{solIso})
and (\ref{nonlocalortotropo}), when a specific orthotropic matrix is considered.
With reference to orthotropic properties of olivine,  pine wood, olivinite, marble, and canine femora
(which orthotropic constitutive parameters are reported in Tab. \ref{tabOrth0} for the three possible orientations
 of orthotropy) used as matrix material,
the corresponding non-null higher-order constants  are given in Tab. \ref{tabOrth} for a dilute suspension of
cylindrical holes with centers aligned parallel to the in-plane
orthotropy axes. All the three possible orientations (Or1, Or2, Or3) are considered
for the axis of the cylindrical inclusion, defining the out-of-plane direction in the plane strain  problem considered.
%%%%%%%%%%%%%%%%%%%%%%%%%%%%%%%%%%%%%%%%%%%%%%%%%%%%%%%%%%%%%%%%%%%%%%
\begin{table}[!htcb]
\centering
\renewcommand{\arraystretch}{1.2}
\renewcommand{\tablename}{\footnotesize{Tab.}}
    \begin{small}
    \begin{tabular}{|c|c|c|c|c|c|c|c|c|c|c|c|c|}
    \hline

\multirow{3}{*}{Matrix material}
&\multirow{3}{*}{Orientation}
&\multirow{3}{*}{$\lambda_1$}
&\multirow{3}{*}{$\mu_1$}
&\multirow{3}{*}{$\xi_1$}
&\multirow{3}{*}{$\omega_1$}  \\[10 mm]

    \hline

\multirow{3}{*}{Olivine}
                            &Or1
                            &$66.000$ & $47.000$ & $-17.000$ & $32.000$\\
                            &Or2
                            &$60.000$ & $106.000$ & $-75.000$ & $-80.000$\\
                            &Or3
                            &$56.000$ & $52.000$ & $-27.500$ & $112.000$\\
\hline
\multirow{3}{*}{Pine (softwood)}
                            &Or1
        &$0.740$ & $8.180$ & $-7.590$ & $-15.860$\\
                            &Or2
        &$0.760$ & $0.515$ & $-0.476$ & $-0.550$\\
                            &Or3
                 &$0.940$ & $8.080$ & $-7.625$ & $-15.310$\\
\hline
\multirow{3}{*}{Olivinite}
                            &Or1
            &$93.000$ & $58.500$ & $-21.85$ & $22.000$\\
                            &Or2
                            &$92.000$ & $53.500$ & $-18.05$ & $33.000$\\
                            &Or3
                            &$82.000$ & $64.000$ & $-29.7$ & $-11.000$\\
\hline
\multirow{3}{*}{Marble}
                            &Or1
        &$51.000$ & $29.500$ & $-14.65$ & $9.000$\\
                            &Or2
                            &$52.000$ & $26.000$ & $-10.65$ & $15.000$\\
                            &Or3
                            &$47.000$ & $31.500$ & $-15.2$ & $-6.000$\\
\hline
\multirow{3}{*}{Canine femora}
                            &Or1
        &$9.730$ & $6.235$ & $-2.900$ & $-3.200$\\
                            &Or2
                 &$           11.900$ & $8.900$ & $-6.065$ & $-10.700$\\
                            &Or3
                   &$         11.900$ & $5.150$ & $-2.815$ & $7.500$\\
\hline
    \end{tabular}
    \end{small}
\caption{\footnotesize{Values of the elastic constants $\lambda_1,\mu_1,\xi_1,\omega_1$ for different orthotropic materials,
namely: olivine (Chevrot and Browaeys, 2004), pine wood (Yamai, 1957), olivinite,
marble (Aleksandrov, Ryzhove and Belikov, 1968), and canine femora (Cowin and Van Buskirk, 1986).
The reported values are in GPa.}}
\label{tabOrth0}
\end{table}
%%%%%%%%%%%%%%%%%%%%%%%%%%%%%%%%%%%%%%%%%%%%%%%%%%%%%%%%%%
%%%%%%%%%%%%%%%%%%%%%%%%%%%%%%%%%%%%%%%%%%%%%%%%%%%%%%%%%%%%%%%%%%%%%%
\begin{table}[!htcb]
\centering
\renewcommand{\arraystretch}{1.2}
\renewcommand{\tablename}{\footnotesize{Tab.}}
    \begin{small}
    \begin{tabular}{|c|c|c|c|c|c|c|c|c|c|c|c|c|}
    \hline

\multirow{3}{*}{Matrix material}
&\multirow{3}{*}{Orientation}
&\multirow{3}{*}{$\ds\frac{a_2}{f\rho^2\mu_1}$}
&\multirow{3}{*}{$\ds\frac{a_4}{f\rho^2\mu_1}$}
&\multirow{3}{*}{$\ds\frac{a_6}{f\rho^2\mu_1}$}
&\multirow{3}{*}{$\ds\frac{a_9}{f\rho^2\mu_1}$}  \\[10 mm]

\hline
\multirow{3}{*}{Olivine}
                            & Or1  &$2.426$ &$1.661$ &$ 3.077$ &$-1.198$\\
                            & Or2  &$1.133$ &$2.105$ &$-1.014$ &$-1.804$\\
                            & Or3  &$3.254$ &$1.497$ &$0.858$ &$-0.780$\\
\hline
\multirow{3}{*}{Pine wood}
                            & Or1  &$0.269$ &$3.789$ &$-3.754$ &$-3.737$\\
                            & Or2  &$10.297$ &$3.551$ &$-3.268$ &$-3.497$\\
                            & Or3  &$0.142$ &$3.478$ &$-3.455$ &$-3.399$\\
\hline
\multirow{3}{*}{Olivinite}
                            & Or1 &$3.119$ &$1.644$ &$-0.220$ &$-1.045$\\
                            & Or2 &$4.398$ &$1.414$ &$0.804$ &$-0.675$\\
                            & Or3 &$4.011$ &$1.481$ &$0.487$ &$-0.782$\\
\hline
\multirow{3}{*}{Marble}
                            & Or1 &$4.023$ &$1.629$ &$-0.257$ &$-1.068$\\
                            & Or2 &$5.866$ &$1.389$ &$0.823$ &$-0.768$\\
                            & Or3 &$5.080$ &$1.532$ &$0.440$ &$-1.015$\\
\hline
\multirow{3}{*}{Canine femora}
                            & Or1 &$8.279$ &$1.219$ &$2.465$ &$-0.801$\\
                            & Or2 &$4.401$ &$2.110$ &$-1.875$ &$-1.788$\\
                            & Or3 &$4.273$ &$1.660$ &$-0.690$ &$-1.063$\\
\hline
    \end{tabular}
    \end{small}
\caption{\footnotesize{
Higher-order equivalent constants $a_2$,  $a_4=a_5$, $a_6$, and $a_9$,
eqns (\ref{solIso}) and (\ref{nonlocalortotropo}), of the orthotropic SGE
material equivalent to an orthotropic matrix containing a dilute suspension of cylindrical holes,
collinear to three possible orientations of orthotropy.
The constants are made dimensionless through division by parameter $f \rho^2 \mu_1$ and are reported for
different matrices, which orthotropy parameters are given in Tab. \ref{tabOrth0}.}
}
\label{tabOrth}
\end{table}
%%%%%%%%%%%%%%%%%%%%%%%%%%%%%%%%%%%%%%%%%%%%%%%%%%%%%%%%%%

\section{Conclusions}

Assuming Cauchy elastic composites made up of a dilute suspension of inclusions and
an RVE with a spherical ellipsoid of inertia, the equivalent higher-order constitutive behaviour
(of \lq Mindlin type') can be defined in a rigorous way, even for anisotropy of
the constituents and complex shape of the inclusions. Through this procedure a perfect
match of the elastic energies of the RVE and of the equivalent higher-order material is obtained,
for a general class of displacements prescribed on the two respective boundaries.
However, it has been shown that, to achieve a positive definite strain energy of the
equivalent higher-order material, the inclusions have to be less stiff (in a way previously detailed) than the matrix,
a situation already found by Bigoni and Drugan (2007) for Cosserat equivalent materials, which limits
the applicability of the presented results, but explains the interpretation of previous experiments and results showing
nonlocal effects for soft inclusions and \lq anti-micropolar' behaviour for stiff ones (Gauthier, 1982).

\vspace*{5mm} \noindent {\sl Acknowledgments}
M. Bacca gratefully acknowledges financial support from Italian Prin 2009 (prot. 2009XWLFKW-002).
D. Bigoni, F. Dal Corso and D. Veber  gratefully acknowledge financial support from the grant PIAP-GA-2011-286110-INTERCER2,
\lq Modelling and optimal design of ceramic structures with defects and imperfect interfaces'.
\vspace*{10mm}

%\clearpage
 { \singlespace
}

\end{document}